\documentclass[11pt, singlespace]{article}

\usepackage{graphicx}
\usepackage{amssymb}
\usepackage{amsmath}
\usepackage{enumerate}
\usepackage{url}
\usepackage{pdfpages}
\usepackage{hyperref}
\hypersetup{
    colorlinks=true,
    linkcolor=blue,
    filecolor=magenta,      
    urlcolor=cyan,
}

\usepackage{enumitem}
\setlist{noitemsep, topsep=0pt, parsep=0pt, partopsep=0pt, leftmargin=1.8cm, labelindent=1.2cm, labelwidth=\wd1, itemindent=*, labelsep=\dimexpr0.6cm-\wd1}
\setbox1=\hbox{$\bullet$}

\newcommand{\be}{\begin{equation}}
\newcommand{\ee}{\end{equation}}
\newcommand{\n}{\noindent}
\newcommand{\dd}{{\rm d}}
\newcommand{\rtarrow}{\,\,\rightarrow\,\,}

\begin{document}

\title{Scattering of Rod-like Swimmers in Low Reynolds Number Environments }
\author{Kentaro Hoeger and Tristan Ursell \\
Department of Physics, University of Oregon\\ Eugene, OR 97424}
\date{ }
\maketitle

In their search for metabolic resources microbes swim through viscous environments that present physical anisotropies, including steric obstacles across a wide range of sizes. Hydrodynamic forces are known to significantly alter swimmer trajectories near flat and low-curvature surfaces. In this work, we imaged hundreds-of-thousands of high-curvature scattering interactions between swimming bacteria and micro-fabricated pillars with radii from $\sim1$ to $\sim10$ cell lengths. As a function of impact parameter, cell-pillar interactions produced distinct chiral distributions for scattering angle – including unexpected ‘counter-rotator’ trajectories – well-described by a sterics-only model. Our data and model suggest that alteration of swimmer trajectories is subject to distinct mechanisms when interacting with objects of different size; primarily steric for objects below ~10 cell lengths and requiring incorporation of hydrodynamics at larger scales. These alterations in trajectory impact swim dynamics and may affect microbial populations in ways that depend on the shape and placement of obstacles within an environment.



**Introduction**

Microbes inhabit chemically complex and physically anisotropic environments – like wet soils or a mammalian gut – often using self-propulsion to find resources and expand into new territory. In these low Reynolds number settings drag quickly dissipates kinetic energy into heat, such that microbes must continually propel themselves to maintain persistent forward movement, and thus their kinematics conserve neither momentum nor energy. Across multiple length scales, swimming microbes interact with their physical environment in ways that alter their trajectories [1]–[8]. For instance, hydrodynamic forces near surfaces potentiate relevant biological phenomena including cell adhesion [9], biofilm formation [10], [11], or colonization of medical devices like catheters [12], [13]. These physical interactions also present opportunities for influencing the motion of micro-swimmers using micro-fabricated environments [14]–[17], for instance to deflect cells from surfaces [18], to passively concentrate them in certain regions [19], or to enhance their motility via flow [20]. *Sipos et al.* [6] used micro-fabricated pillars to show that when swimming near convex surfaces with sufficiently small curvatures, hydrodynamic forces 'trap' cells in ~2D trajectories within ~1 cell diameter of the surface. Similarly, bacteria propelled by helical (and hence chiral) flagella have been observed to swim in approximately constant-curvature trajectories staying close to flat surfaces for minutes at a time [21], with the direction of trajectory rotation linked to flagellar helicity [22]. However, above a critical curvature, entrapment decreases, and for pillars of radius less than ~50 μm the fraction of trapped cells rapidly decreases to zero [6]. Further, such surface trapping was reduced by collisions with small colloids ($r$ = 1.5 μm), which increased the rate of forward scattering and hence increased trajectory persistence [4].

Current theory describes swimming cells as force dipoles with a surrounding toroidal 'Stokeslet' flow field [23] extending more than 10 μm from the cell surface, thus trajectory alterations that arise from interactions with steric obstacles are thought to be primarily hydrodynamic [24]. Model predictions vary due to differences in physical and geometrical assumptions, but generally reproduce the attractive trapping exhibited by low-curvature surfaces. However, it is unknown

2

whether such models accurately describe alterations in trajectories upon interaction with small obstacles on the order of 1 to 10 cell lengths. These size scales are relevant, for instance, in (wet) soil types where measured particle-size distributions indicate that the majority of particles (by number) with which a swimming microbe will interact are less than 10 cell lengths [25], [26]. Further, experimental results from [4] suggest that interactions with particles in the 1-10 cell length range lead to scattering angles that are significantly smaller than those predicted by hydrodynamic models in similar geometries [24], [27].

Thus, while interactions with high curvature surfaces favor forward scattering, the relative roles of hydrodynamics and sterics in the switch between entrapment and forward scattering remain unclear. In this work, we probed high-curvature scattering by imaging hundreds-of-thousands of interactions between flagellar-propelled fluorescent *Escherichia coli* and convex surfaces with positive curvature between 1 and 10 inverse cell lengths. We fabricated microfluidic devices in which bacteria swam among arrays of vertical pillars ranging in size from $R$ = 3.4 µm to 31 µm. We computationally analyzed images to identify trajectories of individual bacteria and characterized scattering events by their impact parameter with respect to a pillar's center. With that data, we calculated the probability distribution of scattering angles for the range of impact parameters $-R \leq b \leq R$. We did not observe hydrodynamic trapping in this range of convex curvatures, but we did observe forward scattering across all measured impact parameters and radii. Across the range of pillar sizes tested, we found that the measured mean scattering angle, exit angle, and interaction time was in quantitative agreement with a relatively simple, fit-free sterics-only model. Our data support a hybrid sterics-hydrodynamics framework for understanding – and potentially controlling – swimmer-surface interactions. Together with previous work, our results underscore that in real-world environments – like ocean particulates [28], soils [25], [26], or a mammalian gut [29], [30] – micron-scale objects influence microbial motion, with potential effects on navigation and subsequent resource acquisition. Further, a physical understanding of how steric objects alter microbial

3

trajectories presents opportunities to design environments that control and/or affect their movements and resulting population dynamics.

**Results**

As cells navigate through real-world environments, like wet sediments or a mammalian gut, they encounter solid, steric objects that alter their trajectory due to both hydrodynamic and steric forces. We wanted to understand the relative role that steric forces play in altering bacterial trajectories at scales about the length of a cell (3.75 µm) to about 10 cell lengths. We built microfluidic devices that present swimming cells with an array of micro-fabricated steric pillars with sizes ranging from $R$ = 3.4 to 31 µm. Cells were cytoplasmically labeled with GFP and their motion was imaged using a fluorescence microscope (see Methods). We imaged hundreds-of-thousands of interactions between swimming bacteria and these steric pillars. To each trajectory we applied custom object tracking algorithms to measure the impact parameter, **b**. We then calculated the outgoing trajectory vector and compared the angle between incoming and outgoing vectors to calculate the scattering angle $\theta$ (Fig. 1A/B). In Figs. 1, 2, and 3 C-E we show data for $R$ = 8.3 µm; data for other pillar radii are shown in the SI.

*Characteristics of cell scattering from high curvature pillars*

For each range of the impact parameter we classified trajectories as either going clockwise (CW) or counter-clockwise (CCW) around the steric pillar, producing normalized probability distributions for scattering angle. The sum of the CW and CCW scattering angle distributions for $R$ = 8.3 µm is shown in Fig. 1C. These distributions exhibited common characteristics across all measured pillar radii. The two 'lobes' of the probability map are produced by the two chiral directions of motion, with the majority of CW paths corresponding to positive impact parameter and the majority

4

4

of CCW paths corresponding to negative impact parameter (Fig. 1B/C). Each lobe has a negative slope, where increasing the magnitude of **b** monotonically decreases the mean scattering angle, which is almost always acute. As the impact parameter approaches the object size ($|b| \to R$) the mean scattering angle approaches, but does not cross, zero, consistent with a lack of hydrodynamic trapping. We are not aware of previous hydrodynamic models that examine our exact scattering geometry (i.e. an upright cylinder between two large, flat surfaces), however, hydrodynamic models of force-dipole swimmers interacting with spherical obstacles (i.e. a similar geometry and similar length scale) [24] predict a wider range of angular deflections that cross $\theta = 0$. Further, we observed that the maximum mean scattering angle increased with pillar radius (SI Fig. 5), which is in contrast to hydrodynamic models that predict longer interaction times, and hence smaller scattering angles for increasing radius of curvature [24]. Likewise, the observed behavior differs from the longer interaction times during surface trapping around flat or low curvature objects [6].

Consistent with previous measurements [4], the vast majority of steric interactions led to forward scattering ($-\pi/2 < \theta < \pi/2$). However, when **b/R** was positive we measured a significant (minority) fraction of swimmers that rotate CCW, extending the CCW-rotator distribution beyond the **b** = 0 centerline, with that fraction decaying to 0 as $b/R \to 1$ (Figs. 1C and 2). Similarly, when **b/R** was negative we measured the same effect mirrored across the **b** = 0 and $\theta = 0$ lines. In either case, we refer to these as 'counter-rotator' trajectories – these forward scattering events correspond to trajectories that traverse the pillar the 'long way' around. We discuss a potential mechanism underlying this effect in the modeling section below.

In Figure 2, we examine the trajectories and chiral angle distributions for three distinct ranges of the impact parameter, again for **R** = 8.3 µm. The top row of Fig. 2 shows the probability that a trajectory passed through a given *XY* position (pixel) – in any *Z*-plane – during a scattering interaction, with CW trajectories shown in green and CCW trajectories shown in magenta. These scattering maps are formed by setting a pixel to one if a trajectory passes through it (zero otherwise) and then

5

averaging over all such binary images for the given range of **b**. This visualizes the general trend between $b/R$ and $\theta$, the statistical nature of these scattering events (i.e. the 'spray' of trajectories that result from distinct ranges of $b/R$), and the spatial distribution of counter-rotator trajectories.

For each chiral direction within a narrow bin of $b/R$ the observed scattering angles were well described by a von Mises distribution with a constant offset (Fig. 2 bottom row). The offset accounts for the small fraction of cells whose interactions with a pillar lead to a uniform, random scattering angle about the unit circle, referred to in [4] as 'tumble-collisions' (see SI Fig. 8). We used maximum-likelihood estimation to fit the mean, width, and offset parameters for these scattering angle distributions as a function of $b/R$, and to determine confidence bounds for those parameters (e.g. see SI Fig. 9). Those fits are shown as dashed lines in Fig. 2 bottom row. We calculated the fraction of trajectories that scattered CW ($p_{CW}$) as a function of impact parameter, and found a smooth and chirally symmetric transition from majority CCW to majority CW as $b/R$ increased from -1 to +1, with larger radii producing a steeper transition (see SI Fig. 10). In the bottom row of Fig. 2 we show the corresponding values of $p_{CW}$ with 95% confidence bounds.

For each binned range of $b/R$, the width of the scattering distribution was approximately constant (~0.3 radians across all data) and chirally symmetric (e.g. see SI Fig. 11). There are likely multiple factors that contribute to this spread in scattering angle, including: rotational and translational diffusion of the cell as it swims; variations in cell length, shape, and axis-of-propulsion; micro-scale surface roughness; and imaging imprecision. Distinct from those sources, the model described below offers a quantitatively consistent mechanism for the observed spread in scattering angle across **b/R**, that relates to the existence of the counter-rotator trajectories.

*Modeling cell scattering*

6

Hydrodynamic forces are known to significantly alter bacterial trajectories near flat and low-curvature surfaces [6], [22]. We wanted to know if steric forces alone could account for the observed interactions between swimming cells and steric pillars. We developed a model that adheres to the following assumptions: (i) hydrodynamic forces and torques between swimmer and pillar surfaces are negligible, (ii) friction between cell and pillar surfaces is negligible (see SI), (iii) the cell is propelled from the rear by a fixed propulsion force **F**, in-line with its long-axis, (iv) the cell is a thin ($R_{cell}/L \ll 1$) stiff rod of length **L**, measured from the center of its flagellar bundle to the cell tip [6], (v) free-swimming motion has a persistence length much larger than the interaction zone, and (vi) that forces that generate rotation of the cell in the plane of the microfluidic device (**$F_R$**) are due to contact between that cell and the solid pillar. Typical Reynolds numbers for swimming bacteria are $10^{-4} - 10^{-3}$, indicating that a constant propulsion force results in a constant cellular velocity (here measured to be ~23 µm/s from the mean free-swimming speed). We model the drag force on each end of the cell by a spherical Stokes drag with particle radius equal to cell radius (0.5 µm) and fluid viscosity equal to that of water. Those assumptions yield a model (see SI for details) where the angle ($\alpha$) that the cell makes with the surface tangent to the point of cell-tip contact is described by the differential equation

$$\dot{\alpha} = -\cos(\alpha)(\sin(\alpha) + \rho) \quad (1)$$

where $\rho = L/R$ and time is measured in the natural scale $L/\langle v \rangle \simeq 0.16s$. In this model, the interaction ceases when the cell's contact angle reaches tangency with the scattering surface ($\alpha = 0$), which corresponds to some rotation angle $\phi$ around the pillar that directly connects with the scattering angle $\theta$ (Fig. 3A). The rate of rotation ($\dot{\alpha}$) is partly due to the rate at which contact sliding along the curved surface causes the curvature of the surface itself to decrease the angle ($\propto -\rho\cos(\alpha)$), and partly due to the projection of the sliding force parallel to the surface and normal to the axis of the cell, which produces a torque over the length of the cell ($\propto -\cos(\alpha)\sin(\alpha)$) (Fig. 3A). Further, we note that the assumptions of this model apply to other rod-like microswimmers propelled

7

on-axis from the rear, including, for instance, abiotic Janus particles [31]. Likewise, while the differential equation above describes interactions with a convex surface of constant radius, the component of the rate change of $\alpha$ with respect to the surface can be adapted to other convex surfaces (concave surfaces present added complexities).

The cell's initial angle of contact with the pillar is uniquely determined by the impact parameter by $\alpha_o = \cos^{-1}(b/R)$. Using a small initial-contact-angle approximation (see SI) – which matches well with the exact numerical solutions used below – the mean scattering angle is approximated by

$$\langle \theta \rangle = \alpha_o^2 \frac{\rho}{4} - \alpha_o \left(\frac{\rho^2}{2} - 1\right)\left(1 - \frac{\rho}{\alpha_o}\ln\left(1 + \frac{\alpha_o}{\rho}\right)\right). \qquad (2)$$

This model has no fit parameters as cell length is externally known (**L** = 3.75 μm, which accounts for the propulsion force acting from part-way into a typical flagellar bundle [6]), pillar radius is measured from electron microscopy of our microfluidic devices (see SI Fig. 12), the fluid viscosity is that of water, the initial contact angle is directly related to **b**, and the average swim speed is measured with our image analysis (and hence application of the Stokes drag gives the average propulsion force **F**). To ensure fidelity to the exact model, we used these known parameters and numerically solved eqn. 1 for the interaction time, exit angle ($\beta$), and scattering angle ($\theta$) at which the condition $\alpha = 0$ is met. In Fig. 3B, we compare the model predictions for mean scattering angle to measured data across four different radii. In Fig. 3C-E we compare the measured distributions for scattering angle, exit angle, and interaction time to the model predictions for **R** = 8.3 μm (other radii shown in SI Figs. 5 – 7).

Overall we find good quantitative agreement between the sterics-only model and the measured scattering distributions over the range $-R \leq b \leq R$, especially for pillar radii $\lesssim 20$ μm. We note, however, that as the radius of the cell approaches the radius of the pillar (i.e. for $R_{\text{cell}} \sim R$) the model assumption of the cell represented by a *thin* stiff rod breaks down for small impact parameters, leading the model to underestimate the scattering angle for small **b**. Likewise, assuming that a cell

8

impacts the pillar surface with its long axis parallel to the incoming scattering vector (e.g. as drawn in Fig. 1A), the steric model predicts all trajectories with **b** > 0 should go CW, while all trajectories with **b** < 0 should go CCW, or said differently, the fraction of CW trajectories would be an increasing, discontinuous step function at **b** = 0. However, as described earlier the existence of counter-rotator trajectories is in contrast to this prediction. Correspondingly, we did not measure a sharp step-function in the fraction of CW rotators vs. $b/R$, though across all radii that function was chirally symmetric, montonically increasing with $b/R$, and crossed $p_{\text{CW}} = 1/2$ at **b** = 0, as expected for any mechanism that obeys the relevant symmetries (see SI Fig. 10).

Our model assumed that initial contact angle ($\alpha_o$) was strictly determined by $b/R$ (i.e. the cell swims straight after entering the interaction zone). To explain the existence of counter-rotator events, we explored the model's predictions when the initial contact angle was offset by an amount $\Delta\alpha$ (see SI Fig. 13A), corresponding to a non-contact rotation of the cell immediately before impact. This approach was motivated by our imaging data, in which we observed cells whose variations in shape and/or axis-of-propulsion caused precession about the long axis as it swam toward a pillar – this produced an overall persistent path, but a 'wobbling' cell axis. Likewise, chemotactic tumble events within the interaction zone could also produce such rotations. Lacking specific knowledge about the distribution of $\Delta\alpha$ (our imaging cannot reliably resolve this momentary shift in orientation), we made the simplest assumption – that $\Delta\alpha$ was a flat distribution, symmetric about zero with a single parameter specifying its width. We chose an angular width of 0.5 radians, or about 1 cell diameter rotating about the cell's center ($L/2$), in either direction. The model predictions for evenly distributed values of $\Delta\alpha$ are shown in SI Fig. 13B for **R** = 8.3 µm. Laid over the measured data, the model predicts shifts in the chiral discontinuity point in $b/R$, a distribution of scattering angles that changes with $b/R$, and concentrations of counter-rotator trajectories that are all consistent with the measured data.

9

**Scattering characteristics for lower curvature pillars**

For larger pillar sizes ($R \geq 20$ μm), the general trend between impact parameter and scattering angle is well-described by the model, but the model consistently overestimates the mean scattering angle (SI Fig. 14). Our model does not easily account for this effect, but these trends are consistent with scattering interactions from lower curvature surfaces being subject to increased hydrodynamic torque that 'over rotates' the cell during the interaction relative to the steric model. For instance, it was shown previously that when cells interact with pillars of radii 20 to 30 μm, hydrodynamic coupling causes a significant fraction of cells (~20%) to be trapped in trajectories that go around the pillar, with interaction times ten-fold longer than our measured interaction times [6]. Thus our data and model are consistent with the hypothesis that for cells scattering from pillars whose radius of curvature is 1 to 10 cell lengths and whose cellular geometry meets the thin-rod condition, the forces and torques that govern scattering are primarily steric in origin.

**Discussion**

We measured a primarily forward-scattering interaction between swimming *E. coli* and surfaces with radii of curvature comparable to cell length and generally larger than cell radius, to determine the relative importance of sterics at these length scales. Various aspects of the mean behavior deviate significantly from hydrodynamic models of similar situations [24], [27] but are well described by a steric model that excludes cell-surface hydrodynamic coupling. This provides strong evidence that swimmers interacting with small, high curvature surfaces are primarily subject to steric forces and that hydrodynamics do not play a significant role in these situations. We did see a significant deviation from our steric model as pillar radii increased, which supports previous experimental findings that hydrodynamic forces play a significant role in describing the motion of bacteria near larger convex surfaces. Taken together, these data suggest that the question of whether swimmer-

10

surface interactions are governed primarily by sterics or hydrodynamics, is one of length scales rather than absolutes.

The sterics-only model makes the additional prediction that swimmers interacting with negative curvature surfaces (concave and where $L/R < 1$) have a stable non-zero contact $\alpha_c = -\sin^{-1}(\rho)$. This effect might be relevant in related studies of the motion of another rod-like flagellated bacterium *Bacillus subtilis* [32]. In that work, swimming cells were contained within a circular hole of radius *R*, effectively presenting the negative curvature analog of a pillar. Their motions were shown to exhibit stable angular orientations with respect to the surface of the circular hole, as measured by the same angle $\alpha$.

We note that our experimental setup has a number of limitations that cannot be circumvented by straightforward engineering. First, the pillar surface is fully characterized by the radius *R*, but it is also a two-dimensional surface described by two principle curvatures, 0 and 1/*R*. It may be that surfaces whose principal curvatures and/or Gaussian-curvature vary produce distinct scattering behavior, potentially (though not necessarily) still well-described by sterics at these length scales. Second, our microfluidic devices had a depth of ~15 μm which has the advantage of permitting full *Z* imaging. However, it is worth noting that swimmer-surface hydrodynamic effects depend on fluid dimensions, because the Stokeslet that describes the propulsive flow-field of flagellated bacteria [23] extends >10 μm microns from the cell surface. Third, the three-dimensional nature of the device also means that incoming trajectories toward a pillar are not necessarily strictly parallel with the plane of the device. These non-parallel scattering interactions likely contribute to both the width of the scattering distributions and potentially as an additional source of counter-rotators. Finally, swimming bacteria are known to exhibit hydrodynamically coupled, chiral motion on surfaces [22]; a small fraction of trajectories exhibited this surface-coupled behavior but not in sufficient numbers to influence scattering statistics.

11

Interaction times from the steric model agree better with measurements as $|b/R| \to 1$; near **b** = 0, however, the initial contact angle approaches $|\pi/2|$ where the model predicts zero net torque on the cell, resulting in very long interaction times. While the data does not show this spike in interaction times about **b** = 0, this is not surprising because both simple rotational diffusion and/or non-zero offset impact angles ($\Delta\alpha$) remove the portion of the contact trajectory that takes the longest, thus the model tends to overestimate the average interaction time near **b** = 0.

Overall the strong, measured correlation between impact parameter and scattering angle suggests that – regardless of the mechanism – the placement of pillars or other steric objects could be used to alter transport properties of cells that are associated with their trajectory [20]; for instance, net directionality, spatial concentration, path persistence length, or mean-squared displacement. Thus, by choosing appropriate values for object size, shape, and position, cells may exhibit distinct patterns of trajectories through arrays of steric objects, allowing experimentalists to influence biologically important aspects of cell motion through the design of micro-fabricated environments.

**Methods**

Experiments used wild-type *Escherichia coli* (HMMG 1655 parent strain) labeled with cytoplasmic monomeric super-folder green fluorescent protein (GFP) under kanamycin selection. Cells were grown from frozen stock in Luria broth with 50 µg/mL kanamycin for 4 hours at 37°C. In order to control chemical inputs to cellular motility and decrease auto-fluorescence of the media, 25 µL of the liquid culture were diluted into 500 µL of a defined minimal media composed of 10 µM thiamine, 100 mM galactose, and 1 mM each of methionine, threonine, and leucine, in a buffer composed of 0.79 mM magnesium chloride, 45 µM Calcium Chloride, 12 µM Ferric Nitrate, 0.34 mM sodium citrate, 7.6 mM Ammonium Sulfate, 27 mM potassium phosphate dibasic, and 12.8 mM potassium phosphate monobasic. We adjusted the dilution, and hence cell density (1 cell / ~1000 µm$^2$), so that the majority of interactions were between a single pillar and a single cell. These

12

interactions were imaged in atypical microfluidic devices that supported significantly larger device aspect ratios than are possible in typical soft polydimethylsiloxane (PDMS) lithography devices [33]. Pillars with radii between 3.4 and 31 µm were patterned onto flat silicon surfaces using SU-8 photoresist (Kayaku Advanced Materials Inc.) exposed with a Hoya L-37 long-pass filter (see SI for fabrication details). To cover and seal the device, we cast a ~100 µm layer of PDMS adhered onto a glass slide which was then mechanically compressed onto the patterned substrate for the duration of image acquisition. The increased stiffness of thin PDMS on glass allowed us to create wide, support-free areas in the device while maintaining a thin fluid layer without the risk of device collapse. Devices consisted of a single chamber 8 mm x 6 mm with a depth of ~15 µm. The device surface was divided into six regions, each patterned with a triangular array of pillars of constant radius with $R = 3.4, 5.8, 8.3, 10.6, 20.5$, and 31 µm (see SI Fig. 12), and an open control region without pillars. Each pillar was spaced at least 10 µm edge-to-edge from neighboring pillars to ensure that each interaction was hydrodynamically independent of nearby pillars. The devices were loaded by pulling the diluted suspension of GFP-tagged *E. coli* via a single inlet, single outlet device layout, and subsequently sealing those ports to halt global flow. We imaged bacterial motion at 21.5 frames per second, for 5 – 10 minutes at a time, with an automated Nikon Eclipse TI-E fluorescence microscope using a Plan Fluor ELWD 20x Ph1 ADM objective and an Andor iXon EMCCD camera. This ultrasensitive camera allowed us to capture images with sufficient signal-to-noise at low illuminations, thus minimizing phototoxic effects on cell physiology and motion. The depth-of-field of the 20x objective allowed us to image cells across the entire *Z*-range of the device. The chamber height constrained cells to move primarily in two dimensions, and thus we did not track vertical motion. Cell segmentation and subsequent *XY* motion tracking were performed by applying a background-subtracted, standard deviation threshold to identify contiguous pixel blocks and their centroids that corresponded to cells. Around each pillar we defined a zone of fixed width ($\delta$ = 2.2 µm); entry of a cell centroid into that zone defined the 'start' of an interaction and exit from that zone defined the 'end' of an interaction. Valid trajectories (those used in the Results) had 10 *XY* positions (10 time

13

points, ~0.5 s) before entering the interaction zone – these points were used to calculate the impact parameter **b** – and 10 *XY* positions after exiting the interaction used to calculate exit angle $\beta$ and scattering angle $\theta$ (Fig. 1). Trajectories were further filtered to exclude cases where: (i) more than one cell was in the interaction zone during the duration of an interaction, (ii) the interaction duration was greater than a cutoff (indicative of possible surface adhesion or physiological issues), or (iii) the trajectory was highly erratic upon entry or exit – defined by a threshold in the absolute curvature of the path immediately before entry or immediately after exit, respectively. We collected between ~30,000 and 100,000 valid interactions per pillar radius yielding a roughly even distribution of sampled impact parameters across $-R < b < R$.

We validated the entire data acquisition and processing pipeline by measuring trajectories of cells in open regions of the microfluidic device devoid of pillars. We defined fictitious pillars ($R_{\text{fict}}$) and interaction zones and applied our image processing to the motion of cells through those regions. Across the range of impact parameters $-R_{\text{fict}} < b < R_{\text{fict}}$ we measured the distributions of scattering angle, exit angle, and interaction duration for bacterial trajectories, subject to the same filtering requirements discussed above. We then compared the means of the scattering angle, exit angle, and interaction duration to a fit-free free-swimming model of dynamics through the fictitious circular interaction zone, with quantitative agreement between the two (SI Fig.15).

**Acknowledgments**

14

**Data Accessibility**

Cell scattering data and scripts are available on the publisher website, and raw imaging data is available upon request.



\begin{figure}
\begin{center}
\includegraphics[width=7in]{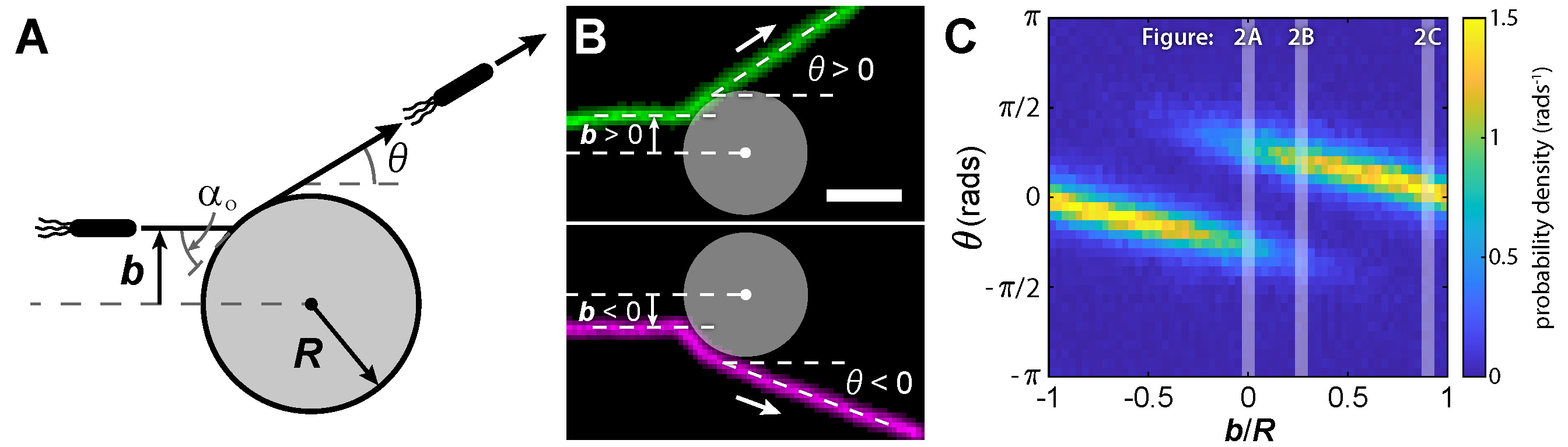}
\caption{Scattering is a chiral and probabilistic process. (A) Schematic showing the impact parameter $b$ for a cell approaching pillar of radius $R$ at an angle $\alpha_o$ and then scattering from the pillar with an outgoing angle $\theta$. As it slides along the pillar surface, the cell rotates and leaves contact with the pillar when its direction of motion, characterized by $\alpha$, is tangent with the pillar surface, leading to a scattering angle $\theta$. (B) Examples of maximum intensity projections of bacterial trajectories interacting with a pillar (drawn in grey) for clockwise (CW) (green) and counter-clockwise (CCW) (magenta) paths. The arrows indicate the direction of movement and the scale bar is $10\mu m$. (C) Heat map showing probability density per radian of an interaction yielding a scattering angle $\theta$ for a given dimensionless impact parameter ($b/R$), here $R = 8.3\,\mu m$. Each column is a normalized distribution. Cells with positive impact parameter tend to slide around the pillar CW leading to a positive scattering angle (right lobe), while cells with negative impact parameter tend to slide CCW leading to a negative scattering angle (left lobe). For each lobe, a minority fraction of trajectories traverses the pillar with the ‘opposite’ handedness (e.g. right lobe for $b/R < 0$). Fig. 2 examines the scattering distributions for the indicated values of $b/R$ (light vertical bars). }
\label{fig1}
\end{center}
\end{figure}

\begin{figure}
\begin{center}
\includegraphics[width=7in]{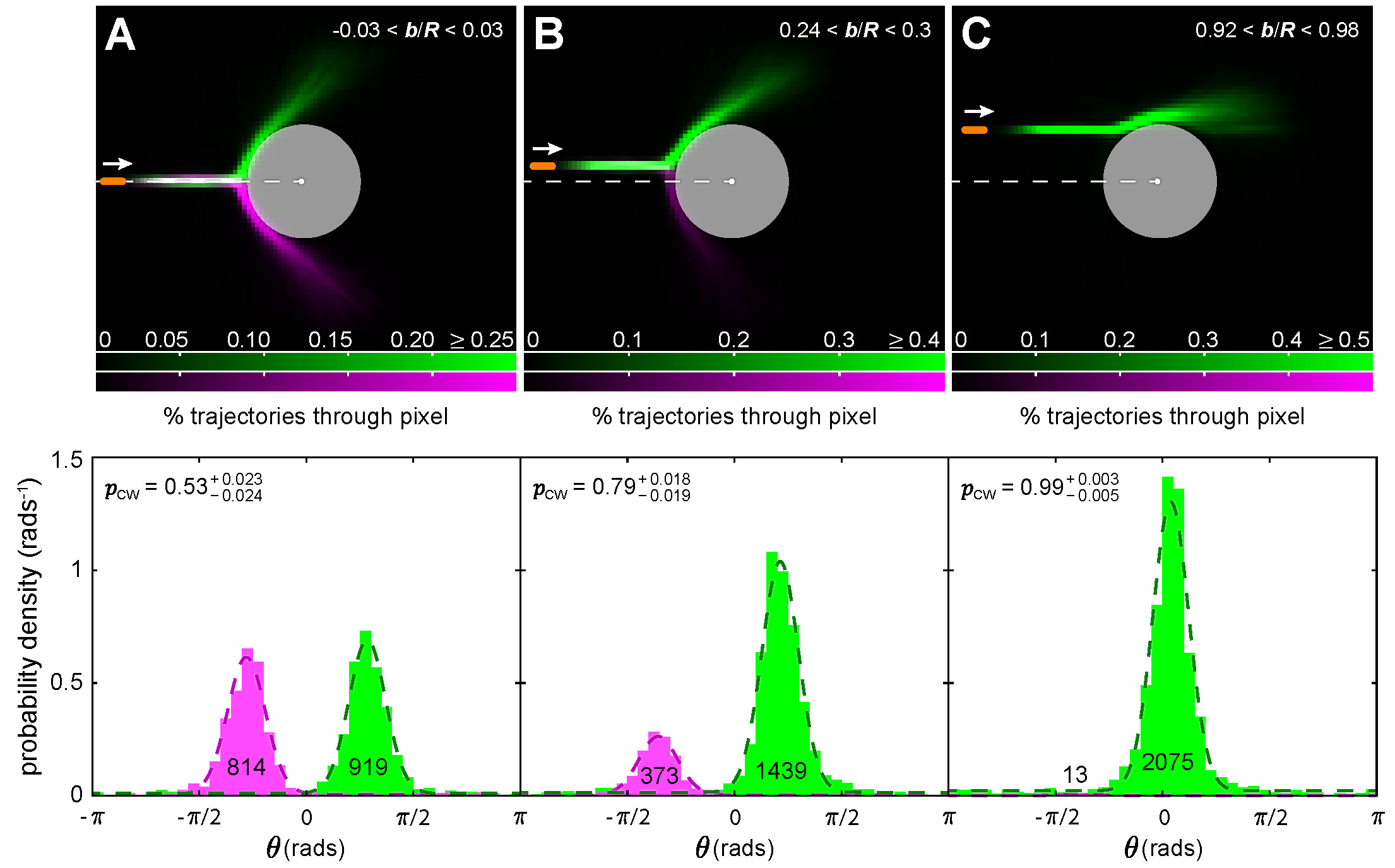}
\caption{Visualizing the statistics of scattering. Each column (A, B, C) shows the aggregated scattering data for $\sim2000$ interactions, for different ranges of $b/R$ as indicated on the top row. (top row) Aligned interaction trajectories for a bacterium (shown approximately to-scale in orange) scattering from a pillar with $R = 8.3\,\mu m$. Green trajectories and histograms correspond to CW paths and magenta trajectories and histograms correspond to CCW paths. In the top row, the color intensity reports on the fraction of trajectories that passed through a given pixel; color saturation was chosen to show a maximum fraction of all trajectories. (bottom row) Each plot shows the normalized distribution for CW (green) and CCW (magenta) scattering angles, with the number of trajectories written on each distribution.  The MLE fits to a modified von Mises distribution are shown as the dashed lines, with corresponding CW probabilities ($p_{CW}$) and 95\% confidence intervals shown in each plot. In general, as $b/R \rightarrow1$, $p_{CW}\rightarrow1$ and $\left<\theta\right>\rightarrow0$.}
\label{fig2}
\end{center}
\end{figure}

\begin{figure}
\begin{center}
\includegraphics[width=7in]{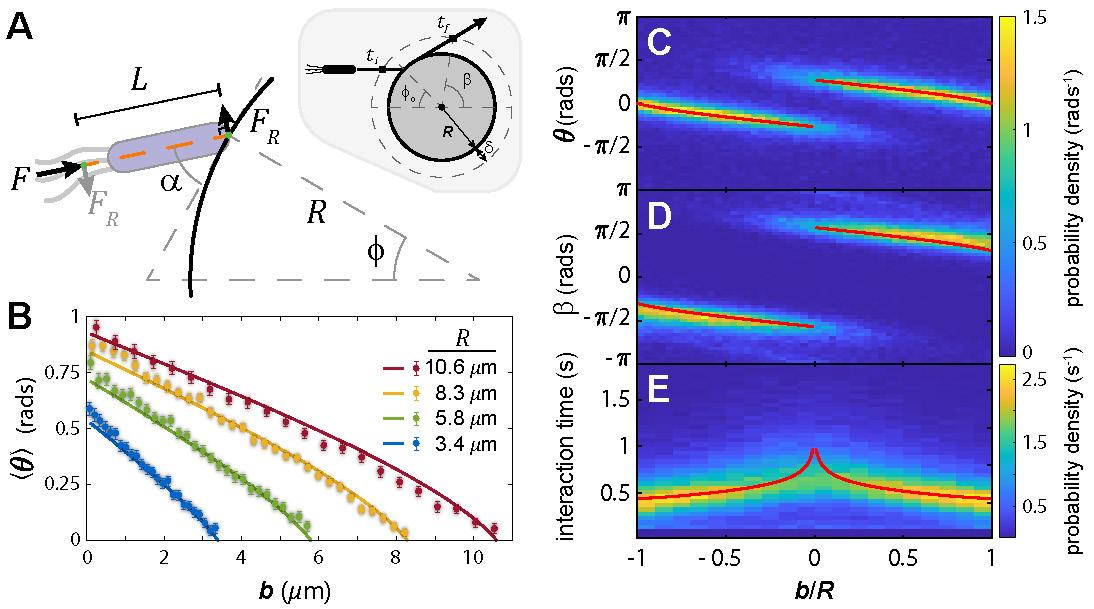}
\caption{Comparing the steric model to experiments. (A) A schematic representation of the forces and geometrical factors described by the sterics-only model. The propulsion force $F$ is generated by the rotation of helical flagella. The length $L$ is the distance – from force-center to cell tip – of the straight, stiff 1D element that $F$ acts on in tangent (orange dashed line). The initial contact angle ($\alpha_o$) is found from the impact parameter $b$. The model assumes that when the angle $\alpha\rightarrow0$ the cell ceases to interact with the pillar. The inset schematic shows the relationship between a trajectory and its exit angle $\beta$, as well as the interaction time, $t_f-t_i$. (B) $\left<\theta\right>$ vs. $b$ extracted from MLE fits of the Von Mises distributions with 95\% confidence intervals, plotted on top of the sterics-only model predictions with $L = 3.75\,\mu m$. (C - E) Scattering angle ($\theta$), exit angle ($\beta$), and interaction time distributions as a function of dimensionless impact parameter $b/R$, with $R = 8.3\,\mu m$. The red lines show the model predictions for the respective measurables. }
\label{fig3}
\end{center}
\end{figure}

\clearpage

\begin{center}
{\LARGE Supplementary Information:\\
Scattering of Rod-like Swimmers in Low Reynolds Number Environments }
\end{center}

\section{Model of Steric Scattering}

Herein we develop a steric model of a rod-like swimmer (e.g.~bacterium) that aligns with a surface and subsequently scatters from it.  Based on observed data, these geometric relationships are sufficient to describe the interaction and the resulting relationship between cellular motion with respect to an oriented surface, specifically predicting the relationship between the scattering parameter $b$ and the mean outgoing angle $\left<\theta\right>$, as well as the duration of interaction (at constant swimming speed) and the angle of exit, $\beta$. Please see the main text for model assumptions.

\subsection{Geometric Constraints}

As a matter of temporary convenience, we assume that the red point in Fig.~1 is the origin of a Cartesian coordinate system.  The motion of each of the points $P_1$ and $P_2$ are parametrically described by $(x_1(t),y_1(t))$ and $(x_2(t),y_2(t))$, respectively, thus all possible dynamics are captured by these four dependent variables. First, note that we are treating the cell as a line-object propelled on-axis from the rear.  We assume that the length of the cell $L$ does not change, mandating that
\be{
(x_2-x_1)^2 + (y_2-y_1)^2 = L^2
}\ee
and we assume (for now) that the point of contact $P_2$ is always in contact, sliding along the surface, until such time as the bacterium leaves the surface, hence
\be{
y_2 =x_2\tan(\theta).
}\ee
The length $L$ is the distance between the leading tip of the cell and the effective point of propulsion, a little longer than the cell body, we use $L=3.75\,\mu m$ throughout this work.

\subsection{Drag-limited Dynamics}

\n
We first build up a simpler model of a swimming cell scattering from a flat surface oriented by an angle $\theta$ with respect to the horizontal (see Fig.~1), and then extend this model to account for movement along a curved (in this case circular) surface of radius $R$.\\

\begin{figure}
\begin{center}
\includegraphics[width=4in]{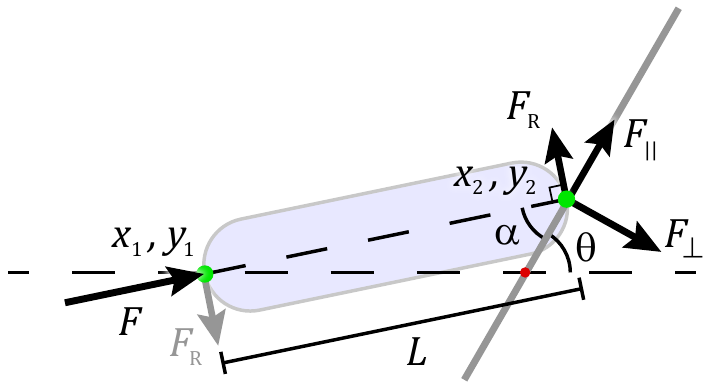}
\caption{Relationships between bacterial orientation ($\alpha$), surface orientation ($\theta$), cell length ($L$), and propulsion force ($F$), for a cell orienting to a flat inclined surface.}
\label{fig1}
\end{center}
\end{figure}

\n
Knowing that swimming bacteria exist at low Reynolds number ($\sim10^{-4} - 10^{-3}$  ), we assume that viscous drag limits movement of the points $P_1$ and $P_2$, and hence that the velocities of points $P_1$ and $P_2$ are proportional to the net force on those points with a fixed mobility $\sigma$ for each point.  The propulsion force $F$, independent of any state of motion can be decomposed into a component that is parallel to the scattering surface $F_{\|}$ and a component normal to the surface $F_{\perp}$, such that given the current angle $\alpha$, 
\be{
F_{\perp}=F\sin(\alpha)
}\ee
and
\be{
F_{\|}=F\cos(\alpha).
}\ee
We approach the equations of motion as a problem of finding $(x_i,y_i)$ as functions of $\alpha(t)$ and its derivatives.  The force parallel to the surface translates the point $P_2$ according to
\be{
\dot{x}_2 = F\sigma \cos(\alpha)\cos(\theta)
}\ee
\be{
\dot{y}_2 = F\sigma \cos(\alpha)\sin(\theta)
}\ee
The distance $x_2-x_1$ is also defined geometrically by
\be{
x_2 -x_1  = L\cos\left(\theta-\alpha\right)
}\ee
and hence its time derivative is
\be{
\dot{x}_2 - \dot{x}_1 = \dot{\alpha} L\sin(\theta-\alpha)
}\ee
such that
\be{
\dot{x}_1 = \dot{x}_2 - \dot{\alpha} L \sin(\theta - \alpha) =  F\sigma \cos(\alpha)\cos(\theta) - \dot{\alpha} L \sin(\theta - \alpha)
}\ee
Looking back at the constraint for $L$ and taking the time derivative
\be{
(x_2-x_1)^2 + (y_2-y_1)^2 = L^2 \rtarrow   (x_2 -x_1)(\dot{x}_2 - \dot{x}_1) + (y_2 -y_1)(\dot{y}_2 - \dot{y}_1) =0
}\ee
Then using our results above
\be{
\dot{\alpha} L\sin(\theta-\alpha) + \frac{y_2 -y_1}{x_2 -x_1}(\dot{y}_2 - \dot{y}_1) =0
}\ee
and with
\be{
 \frac{y_2 -y_1}{x_2 -x_1}= \tan(\theta-\alpha)
}\ee
this simplifies to
\be{
\dot{\alpha} L\cos(\theta-\alpha) + \dot{y}_2 - \dot{y}_1 =0 \rtarrow \dot{y}_1 = \dot{y}_2 + \dot{\alpha} L\cos(\theta-\alpha)
}\ee
and finally
\be{
\dot{y}_1 = F\sigma \cos(\alpha)\sin(\theta) + \dot{\alpha} L\cos(\theta-\alpha)
}\ee
Then the projection of the translational force $F_{\|}$ onto the coordinate perpendicular to the axis of the cell is what causes the cell body to rotate with respect to the surface, and thus
\be{
F_R=F_{\|}\cos\left(\frac{\pi}{2}-\alpha\right) = F\sin(\alpha)\cos(\alpha)
}\ee
Finally, rotation of the cell is
\be{
\dot{\alpha}  - \frac{F_R\sigma}{L} = - \frac{F\sigma}{L}\sin(\alpha)\cos(\alpha). 
}\ee\\

\n
We note that the natural length scale is $L$ (as it has nothing to do with $R$) and the natural time scale is $L/F\sigma$, such that the equations of motion can be non-dimensionalized and written
\be{
\dot{\alpha} = - \sin(\alpha)\cos(\alpha)
}\ee
and then
\be{
\dot{x}_2 = \cos(\alpha)\cos(\theta)
}\ee
\be{
\dot{y}_2 = \cos(\alpha)\sin(\theta)
}\ee
\be{
\dot{x}_1 =  \cos(\alpha)\cos(\theta) - \dot{\alpha} \sin(\theta - \alpha)
}\ee
\be{
\dot{y}_1 = \cos(\alpha)\sin(\theta) + \dot{\alpha} \cos(\theta-\alpha)
}\ee
Finally, the differential equation for $\alpha$ with initial condition $\alpha(0) = \alpha_o$ is solved by
\be{
\alpha(t) = -\frac{1}{2}\tan^{-1}\left[\frac{2e^{-t}\tan(\alpha_o)}{1+(e^{-t}\tan(\alpha_o))^2}\mbox{ ,  }\frac{1-(e^{-t}\tan(\alpha_o))^2 }{1+(e^{-t}\tan(\alpha_o))^2}\right]
}\ee
where the effect of the initial condition is to shift the time axis by $t_o = -\ln(\tan(\alpha_o))$. For long times or small $\alpha_o$ this can be approximated simply as
\be{
\alpha(t) \simeq \alpha_o e^{-t}.
}\ee
This was the case for a rod-like object orienting to a flat surface tilted by an angle $\theta$.

\subsection{Contact Friction}

To determine the potential role of friction, we note that if the parallel force exceeds the friction force then the point of contact will move, this can be stated as
\be{
F_{\|}\geq \mu F_{\perp}
}\ee
where $\mu$ is the frictional coefficient, which gives a critical impact angle of
\be{
\alpha_c = \tan^{-1}\left(\frac{1}{\mu}\right).
}\ee
This is a condition for the balance between frictional and sliding forces -- our data frequently show cells impacting the steric object essentially head-on, with subsequent sliding along the surface, indicating that the friction $\mu\ll 1$, supporting the model assumption that the motion is drag-limited.

\begin{figure}
\begin{center}
\includegraphics[width=3in]{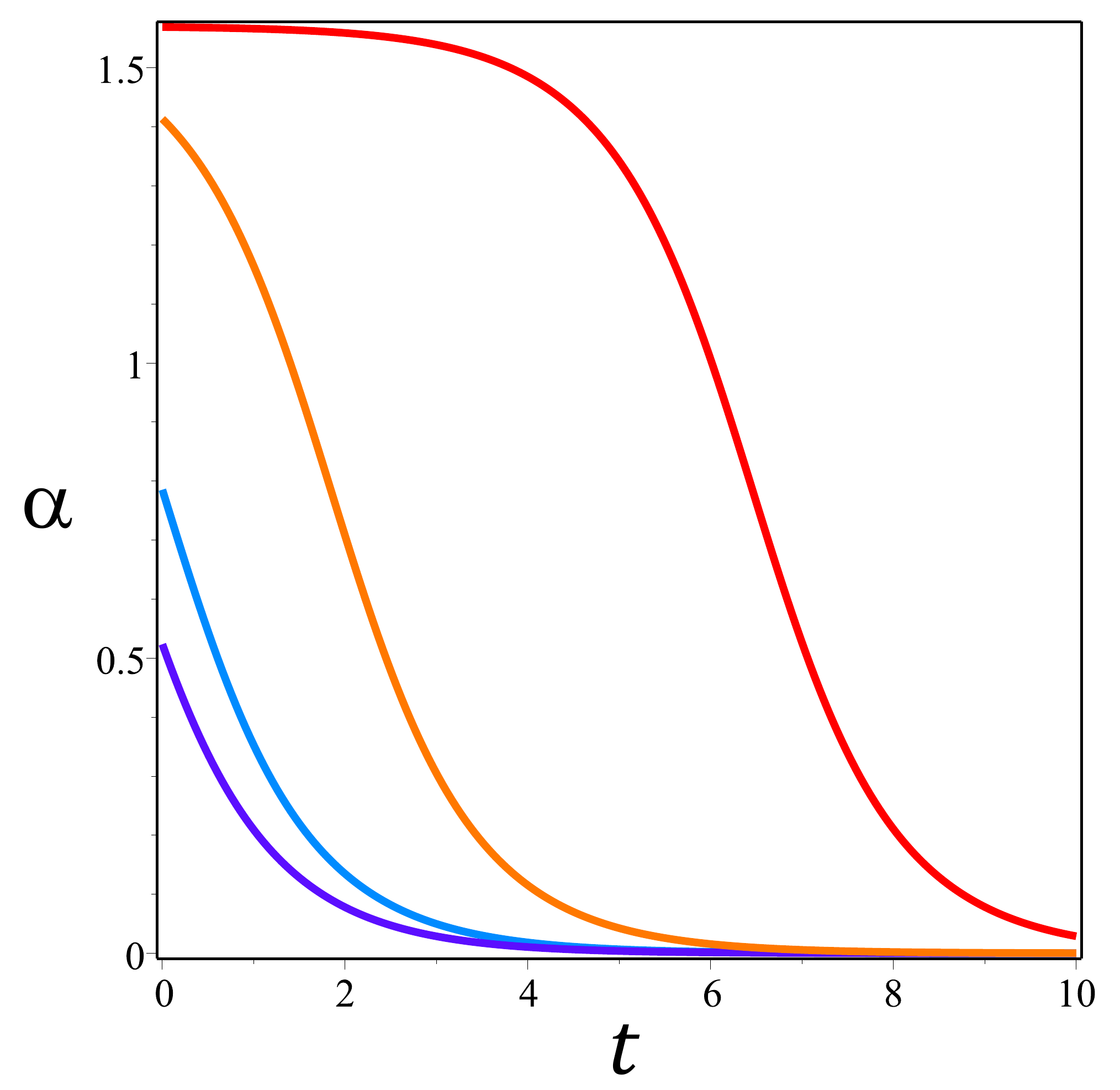}
\caption{Bacterial orientation ($\alpha$) with respect to a flat inclined surface as a function of time ($t$) in dimensionless units for (left to right) $\alpha_o = \pi/6, \pi/4,0.9\pi/2,0.999\pi/2$.}
\label{fig2}
\end{center}
\end{figure}

\subsection{Interactions with a Curved Surface}

Assuming that viscous drag is the primary constraint on motion, we assume that all velocities are proportional to net force with a fixed mobility $\sigma$.  The propulsion force $F$, independent of any state of motion can be decomposed into a component that is parallel to the scattering surface $F_{\|}$ and a component normal to the surface $F_{\perp}$, such that given the current angle $\alpha$, 
\be{
F_{\perp}=F\sin(\alpha)
}\ee
and
\be{
F_{\|}=F\cos(\alpha).
}\ee
For simplicity assume that the circle's center is the coordinate origin, and hence
\be{
x_2 = -R\cos(\phi)
}\ee
\be{
y_2 = R\sin(\phi)
}\ee
and thus
\be{
\dot{x}_2 = \dot{\phi} R\sin(\phi)
}\ee
\be{
\dot{y}_2 =  \dot{\phi} R\cos(\phi)
}\ee
Using the parallel force we can also write
\be{
\dot{y}_2 = F_{\|} \sigma \cos(\phi)=F\sigma\cos(\alpha)\cos(\phi)
}\ee
\be{
\dot{x}_2 = F_{\|} \sigma \sin(\phi)=F\sigma\cos(\alpha)\sin(\phi)
}\ee
Both of these equations dictate that
\be{
\dot{\phi}=\frac{F\sigma}{R}\cos(\alpha)
}\ee
which using the same definitions of time and length scale give 
\be{
\dot{\phi}=\rho\cos(\alpha)
}\ee
with $\rho = L/R$, and the initial condition is related to the impact parameter by
\be{
\phi_o = \sin^{-1}\left(\frac{b}{R}\right)
}\ee
and likewise the initial value of $\alpha$ is
\be{
\alpha_o = \frac{\pi}{2}-\phi_o
}\ee
because we assume the cell impacts in a flat orientation (i.e.~$y_1 = y_2$).  Then the rate change of $\alpha$ due to {\it torque} is 
\be{
\dot{\alpha}_T = - \frac{F_R\sigma}{L}
}\ee
where
\be{
F_R = F_{\|}\cos\left(\frac{\pi}{2}-\alpha\right) = F_{\|}\sin(\alpha) = F\cos(\alpha)\sin(\alpha)
}\ee
and the rate change of $\alpha$ due to the {\it surface curvature} is
\be{
\dot{\alpha}_C = -\dot{\phi}
}\ee
then
\be{
\dot{\alpha}  = \dot{\alpha}_T + \dot{\alpha}_C = \frac{F\sigma}{L}\cos(\alpha)\sin(\alpha) - \frac{F\sigma}{R}\cos(\alpha)
}\ee
and upon non-dimensionalization
\be{
\dot{\alpha} = -\cos(\alpha)\sin(\alpha) -\rho\cos(\alpha) = -\cos(\alpha)\left(\sin(\alpha)+\rho\right)
}\ee
This model predicts that if the cell is perpendicular to the surface $(\alpha = \pi/2)$ then $\dot{\alpha} = 0$, same as the flat surface.  However, it also predicts that there is a non-zero critical angle 
\be{
\alpha_c = -\sin^{-1}\left(\rho\right) \rtarrow \rho <1
}\ee  
that results in a stable orientation with respect to the surface, however, the fact that that angle is negative means that this only occurs for cells on the `inside' (i.e. negative curvature), which may be part of the consistent orientation of motile {\it Bacillus subtilis} cells observed on the {\it inside} curvature of a circle\footnote{E.~Lushi, H.~Wioland, R.E.~Goldstein; Fluid flows created by swimming bacteria drive self-organization in confined suspensions (2014). {\it PNAS} {\bf 111}, 9733 - 9738.}.

\begin{figure}
\begin{center}
\includegraphics[width=4in]{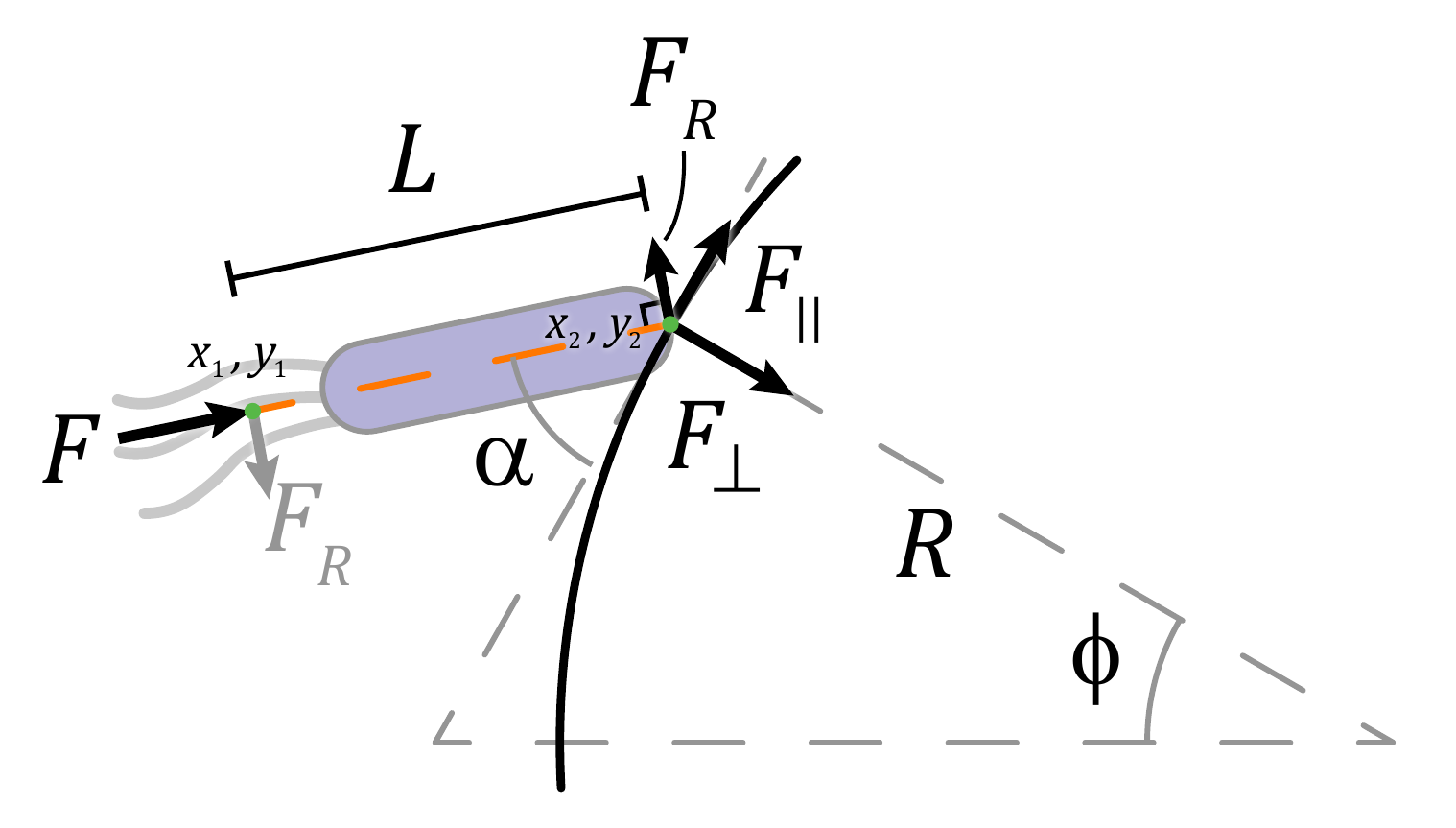}
\caption{Relationships between the various forces and geometrical parameters of the circular model, including bacterial orientation ($\alpha$), surface orientation ($\phi$), cell length ($L$), and propulsion force ($F$).}
\label{fig3}
\end{center}
\end{figure}

For the moment let us make analytic headway by assuming small $\alpha_o$, and thus the differential equation becomes
\be{
\dot{\alpha}\simeq -\alpha -\rho \rtarrow \alpha = e^{-t}\left(\alpha_o + \rho\right) - \rho,
}\ee
noting that the flat surface case (earlier) corresponds to $\rho\rightarrow 0$. The assumption of the model is that the bacterium leaves the surface when $\alpha=0$, thus the time when that happens is
\be{
t_c = \ln\left(\frac{\alpha_o}{\rho} + 1\right)
}\ee
and the angle $\phi$ at which it leaves is determined by
\be{
\dot{\phi}=\rho\cos(\alpha) \rtarrow \phi_c = C + \rho\int_0^{t_c} \cos(\alpha)\dd t \simeq  C + \rho\int_0^{t_c} \left[1-\frac{\alpha^2}{2}\right]\dd t
}\ee
where $C$ is a constant such that $\phi(0)=\phi_o$. This integral has a complicated solution, however approximating cosine by its first two Taylor series terms we can find
\be{
\phi_c = \frac{\pi}{2} -\rho\frac{\alpha_o^2}{4} +\alpha_o\left(1 - \frac{\rho}{\alpha_o}\ln\left(\frac{\alpha_o}{\rho} + 1\right)\right)\left( \frac{\rho^2}{2}-1\right)
}\ee
Then finally, the measured exit angle is given by
\be{
\theta = \frac{\pi}{2} - \phi_c =\rho\frac{\alpha_o^2}{4} -\alpha_o\left(1 - \frac{\rho}{\alpha_o}\ln\left(\frac{\alpha_o}{\rho} + 1\right)\right)\left( \frac{\rho^2}{2}-1\right)
}\ee
with $\alpha_o = \cos^{-1}\left(\frac{b}{R}\right)$. Similarly, the limit when $\rho\rightarrow 0$ gives the initial condition $\theta = \alpha_o$, consistent the flat-surface model.  The models overlaid with data in the main text and SI were calculated using this differential equation, but were solved exactly (numerically) (as opposed to applying the small $\alpha_o$ approximation).

\begin{figure}
\begin{center}
\includegraphics[width=3.5in]{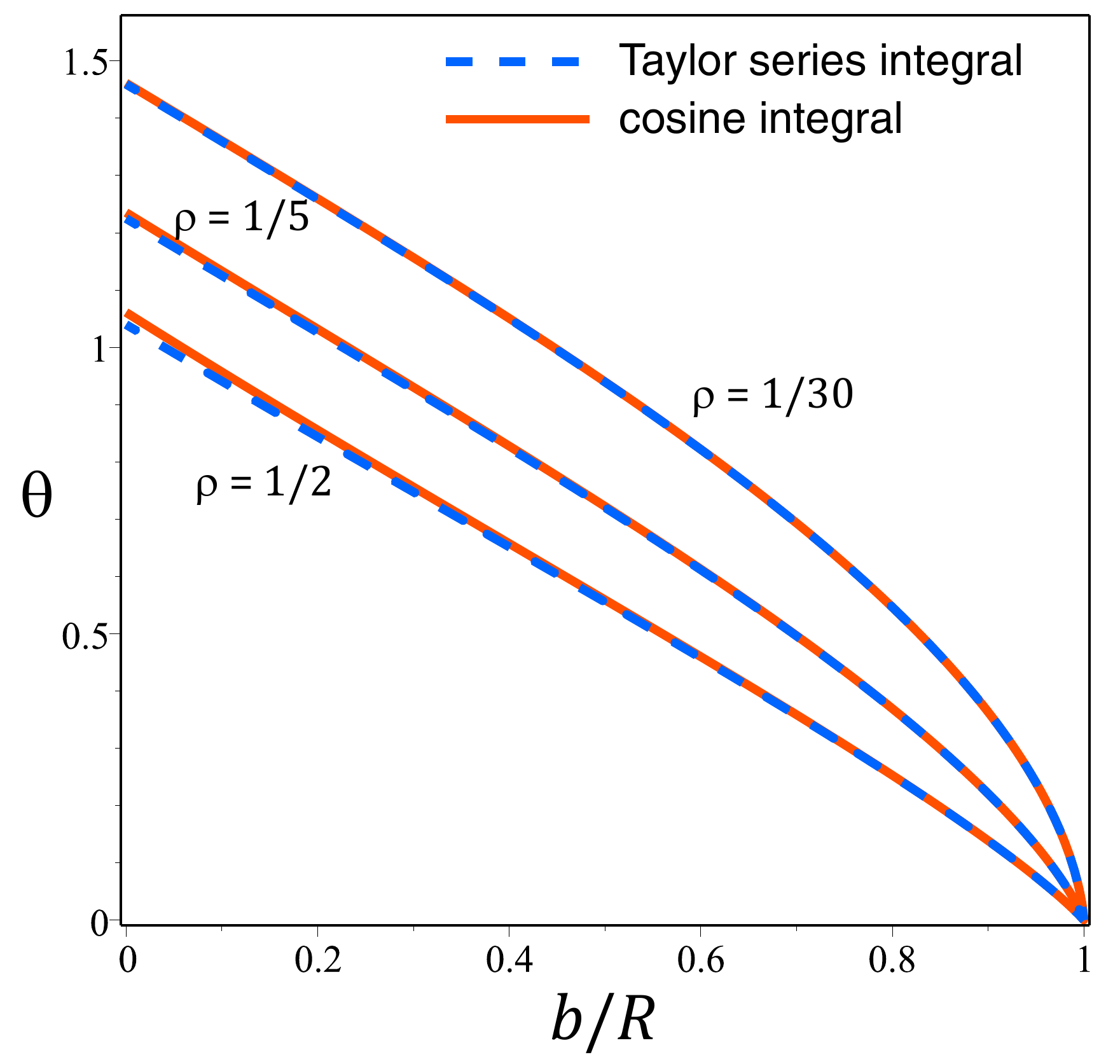}
\caption{Relationship between impact parameter $b/R$ and the output angle $\theta$ for values of $\rho$ indicated.}
\label{fig4}
\end{center}
\end{figure}

\subsection{Interaction Time}

An interaction with a pillar of radius $R$ was computationally triggered when a bacterium came within $R+\delta$ of the pillar center, where $\delta$ is the radial zone around the pillar inside of which we measured interactions, usually $2 - 3\,\mu m$ from the pillar surface. Thus for a given value of $b$, the initial straight line path from entry into the interaction zone until contact with the pillar has a length
\be{
s_1 = R\left[\sqrt{\left(1+\frac{\delta}{R}\right)^2-\left(\frac{b}{R}\right)^2} - \sqrt{1-\left(\frac{b}{R}\right)^2}\right]
}\ee
and applying the average swim speed $\left<v\right>$, a transit time of
\be{
t_1 = \frac{s_1}{\left<v\right>}.
}\ee
Likewise, after the cell has slide around the pillar and rotated to be tangent with the pillar surface, the length from that point to exit of the interaction zone is
\be{
s_3 = R\sqrt{\left(1+\frac{\delta}{R}\right)^2-1}
}\ee
and a transit time of
\be{
t_3 = \frac{s_3}{\left<v\right>}
}\ee
The time spent sliding and rotating around the pillar can be found exactly from the differential equation
\be{
{\dot \alpha}=-\cos(\alpha)(\sin(\alpha)+\rho)
}\ee
which can be integrated directly for the time at which certain values of $\alpha$ are achieved
\be{
t + C = -\int\frac{\dd \alpha}{\cos(\alpha)(\sin(\alpha)+\rho)} = \frac{\ln(\sin(\alpha)+1)}{2-2\rho}+\frac{\ln(\sin(\alpha)-1)}{2+2\rho}+\frac{\ln(\sin(\alpha)+\rho)}{(\rho+1)(\rho-1)}
}\ee
where $C$ is an unimportant constant. The time between contact and tangency is given by
\be{
t_2 = \frac{L}{\left<v\right>}\left[\left.t\right|_{\alpha=0} - \left.t\right|_{\alpha=\alpha_o}\right]
}\ee
where we have now accounted for the natural timescale, and this simplifies to
\be{
t_2 = -\frac{L}{\left<v\right>}\left[\frac{\ln(\sin(\alpha_o)+1)}{2(1-\rho)} + \frac{\ln\left(\left|\sin(\alpha_o)-1\right|\right)}{2(1+\rho)} + \frac{\ln\left(\frac{\sin(\alpha_o)}{\rho}+1\right)}{(\rho-1)(\rho+1)}\right]
}\ee
Then the total interaction time is
\be{
t_{\mbox{\tiny int}}=t_1+t_2+t_3 = t_f - t_i
}\ee
See SI Figure \ref{figinttime}. In our data processing, we subtract a constant length (of $1\,\mu m$) from $s_1$ to account for the offset between the position of the tip which makes contact with the pillar and the position of the cell centroid from image processing, that offset is applied consistently to all data processing and figures. 

\section{Predictions for Control Data}

As a test for our entire image analysis and data pipeline, we imaged cells swimming through open regions of our device, that is, devoid of any steric obstruction except the upper and lower surfaces.  We created fictitious interaction by zones by defining a typical (fictitious) pillar dimension ($R=5.8\,\mu m$) and corresponding interaction zone of width $\delta = 2.2\,\mu m$.  As bacteria swam through the interaction zone, we processed their trajectories in precisely the same way as we processed actual steric interactions.  We constructed the same plots of: dimensionless impact parameter ($b/R$) vs. scattering angle ($\theta$), $b/R$ vs. exit angle ($\phi$), and $b/R$ vs. interaction time, and we calculated the expected mean values of those relationships.  The calculations below assume that the persistence length of the isotropic persistent random walk of the cellular trajectories is much longer than $R+\delta$. 

In particular, if diffusion of a trajectory across the interaction zone was isotropic, then the entry angle (of 0) should, on average, be zero upon exit, regardless of $b$ and hence
\be{
\left<\theta\right>(\tfrac{b}{R}) = 0.
}\ee
Similarly, if diffusion is isotropic the point of entry into the interaction zone, specified by $b$, has the same mean $y$-axis ($y=b$) value at the point of exit, giving the exit angle of 
\be{
\left<\beta\right>=\sin^{-1}\left(\frac{b/R}{1+\frac{\delta}{R}}\right)
}\ee
Finally, the interaction time, that is, the time from entry to exit, will be dominated by  approximately straight trajectories that exit, on average, at the same $y=b$ value at both points.  The time to execute that trajectory is
\be{
t_{\mbox{\tiny int}}=2\frac{R}{\left<v\right>}\sqrt{\left(1+\frac{\delta}{R}\right)^2-\left(\frac{b}{R}\right)^2}.
}\ee
\n
The data and overlaid control models are shown in Fig.~\ref{figcontrol}.

\section{Measured Chiral Symmetry}

Given the mid-plane reflection symmetry of the device (in $Z$) we expected the CW- and CCW-rotator distributions (including counter-rotators) to be approximately symmetric when mirrored across the $b = 0$ and $\theta=0$ lines. We tested this by applying the appropriate symmetry operations to the data and then compared the mean scattering angles of each lobe for $0\leq|b/R|\leq 1$. For each pillar radius the mean scattering angles between the two lobes were largely symmetric. As pillar radius increased, there was a small chiral asymmetry between the two lobes (SI Fig. \ref{fig_chiral_test}). Through initial, iterative improvement of the fabrication process we observed that decreasing the systematic tapering of pillars – resulting from photolithography – reduced these chiral asymmetries. Thus the observed asymmetry likely arises from small, systematic pillar tapering ($\leq4\%$) that asymmetrically affects chiral coupling at the upper and lower surfaces where the difference in pillar radius is greatest.  

\section{MLE Fitting}

In order to extract parameters that both describe the trends of the scattering process and to compare with the predictions of our model, we applied Maximum-likelihood estimation to determine parameter values and 95\% confidence intervals. For each bin in $b$, we started with a von Mises distribution modified to include a constant offset that accounts for the uniform scattering angle that corresponds to non-directional `tumble-collisions' in our measured data
\be{
\rho(\theta;\left<\theta\right>,\sigma,c) = \frac{c}{1+2\pi c}\left(1+\frac{e^{\frac{\cos(\theta-\left<\theta\right>)}{\sigma^2}}}{2\pi c I_0(\sigma^{-2})}\right)
}\ee
where $\theta$ is the measured scattering angle, $\sigma$ is the width of the distribution in radians (analogous to the standard deviation of a Gaussian), $\left<\theta\right>$ is the mean scattering angle, $c$ is the offset parameter, and $I_0$ is the modified Bessel function of the first kind. The index $i$ spans the measured values of $\theta$. The log-likelihood function is then
\be{
\ln(\mathcal{L}(\left<\theta\right>,\sigma,c))=\sum_{i=1}^N\ln(\rho(\theta_i;\left<\theta\right>,\sigma,c))
}\ee
which simplifies to
\be{
\ln(\mathcal{L}) = N\ln\left(\frac{c}{1+2\pi c}\right) + \sum_{i=1}^N\ln\left(1+\frac{e^{\frac{\cos(\theta_i - \left<\theta\right>)}{\sigma^2}}}{2\pi c I_0(\sigma^{-2})}\right)
}\ee
where the fraction of tumble-collisions is
\be{
f_{\mbox{\tiny tumb}}=\frac{2\pi c}{1+2\pi c}
}\ee
We numerically sampled the log-likelihood function over reasonable ranges of all three parameters, and found the mode values for the parameters with 95\% confidence intervals specified from the respective marginal distributions. An example of this data processing routine is shown SI Fig.~\ref{fig_mle_fit}.

\section{Device Fabrication}

Bacterial scattering events were measured in atypical microfluidic devices composed of a silicon wafer patterned with photoresist, and mechanically compressed against a thin layer of PDMS that was bonded to a glass slide. The top of the device consisted of a $5\,cm$ silicon wafer (University Wafer) onto which we spun a $0.5\,\mu m$ base layer of SU-8 2000.5 negative photoresist (Kayaku Advanced Materials Inc.). That layer was first soft baked at $95\,C$ for 1 minute, exposed at an energy density of $60\,mJ/cm^2$, and baked at $95\,C$ for another minute to cure the layer. This base layer increases adhesion of the pillars to the surface and improves feature resolution.  Onto this existing layer of cured photoresist, we spun a $\sim15\,\mu m$ layer of SU-8 2015 negative photoresist, and then soft baked it at $95\,C$ for three minutes. This thicker layer of photoresist was exposed with a quartz chromium mask containing the flow layout and pillared regions within the device, using a Suss MJB4 mask aligner. ‘T-topping’ (i.e.~pillar taper) was minimized by filtering wavelengths below $360\,nm$ using a Hoya L-37 longpass filter (Hoya Optics Inc.) with an exposure energy density of $240\,mJ/cm^2$. The photoresist was developed by mildly agitating the silicon wafer in SU-8 developer for 3 minutes and then performing a final ‘hard bake’ for 10 minutes at $200\,C$ to increase structural stability.

The bottom piece consists of a thin layer of PDMS bonded to a glass slide that has inlet and outlet ports pre-drilled. Uncured PDMS is compressed between the pre-drilled slide and a second glass slide treated with tichlorosilane to minimize adhesion of the PDMS to this second slide. Small adhesive spacers between the two slides fixed the PDMS layer thickness to be $\sim100\,\mu m$. The PDMS was bonded to the drilled slide by baking at $100\,C$ for 90 mins. Excess PDMS was removed from the inlet and outlet ports using a 1 mm biopsy punch. The patterned silicon wafer was then aligned to the inlet and outlet ports and mechanically compressed 
to create an airtight seal suitable for pulling suspensions of cells through the device with a syringe.

Once filled with the cellular suspension, the device ports were sealed 
to halt any global flow, and the device was viewed from the bottom through the glass slide on an inverted microscope.

\section{Supporting Figures}

\begin{figure}
\begin{center}
\includegraphics[width=7in]{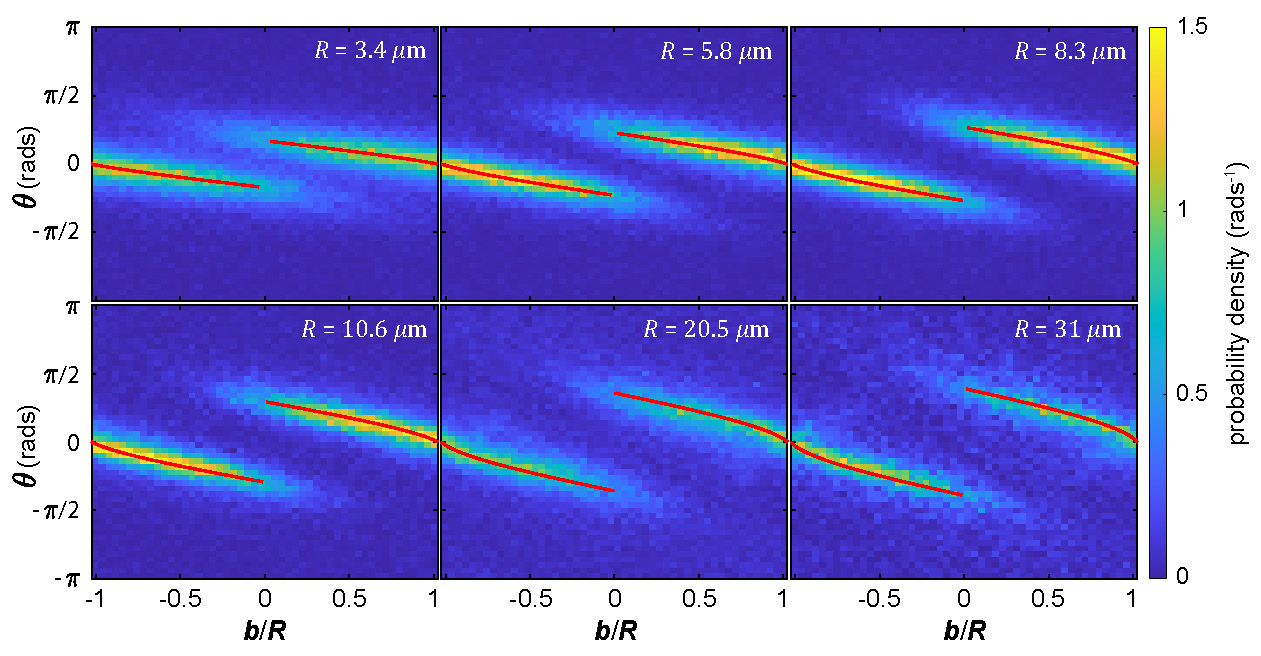}
\caption{Scattering angle distributions as a function of dimensionless impact parameter $b/R$ (same type of data as shown in Fig.~3C) across a range of pillar radii.  The red lines show the model predictions for $\left<\theta\right>$ given the listed radii.  All calculations use the same exogenously specified cell length of $L = 3.75\,\mu m$. Notably, the `signal-to-noise' ratio of measured data decreases with increasing pillar radius because the the number of pillars and hence number of interactions we can observe in a single field-of-view decreases faster than $R^{-2}$.}
\label{fig_theta_all_radii}
\end{center}
\end{figure}

\begin{figure}
\begin{center}
\includegraphics[width=7in]{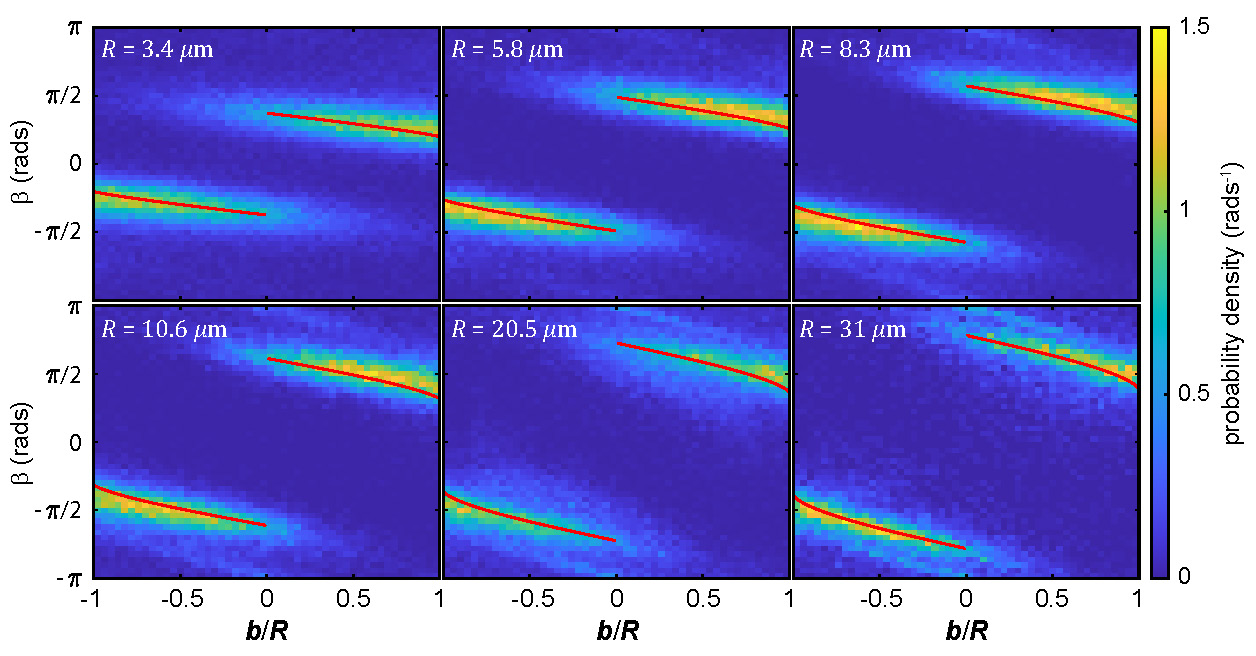}
\caption{Interaction zone exit angle distributions ($\beta$) as a function of dimensionless impact parameter $b/R$, across a range of pillar radii (same type of data as Fig.~3D). The red lines show the model predictions for $\left<\beta\right>$ given the listed radii. Model predictions were calculated by using the first cell trajectory point (in the rotated frame) outside of the interaction radius upon exit. All calculations use the same exogenously specified cell length of $L = 3.75\,\mu m$. Notably, the `signal-to-noise' ratio of measured data decreases with increasing pillar radius because the the number of pillars and hence number of interactions we can observe in a single field-of-view decreases faster than $R^{-2}$.}
\label{fig_beta_all_radii}
\end{center}
\end{figure}

\begin{figure}
\begin{center}
\includegraphics[width=7in]{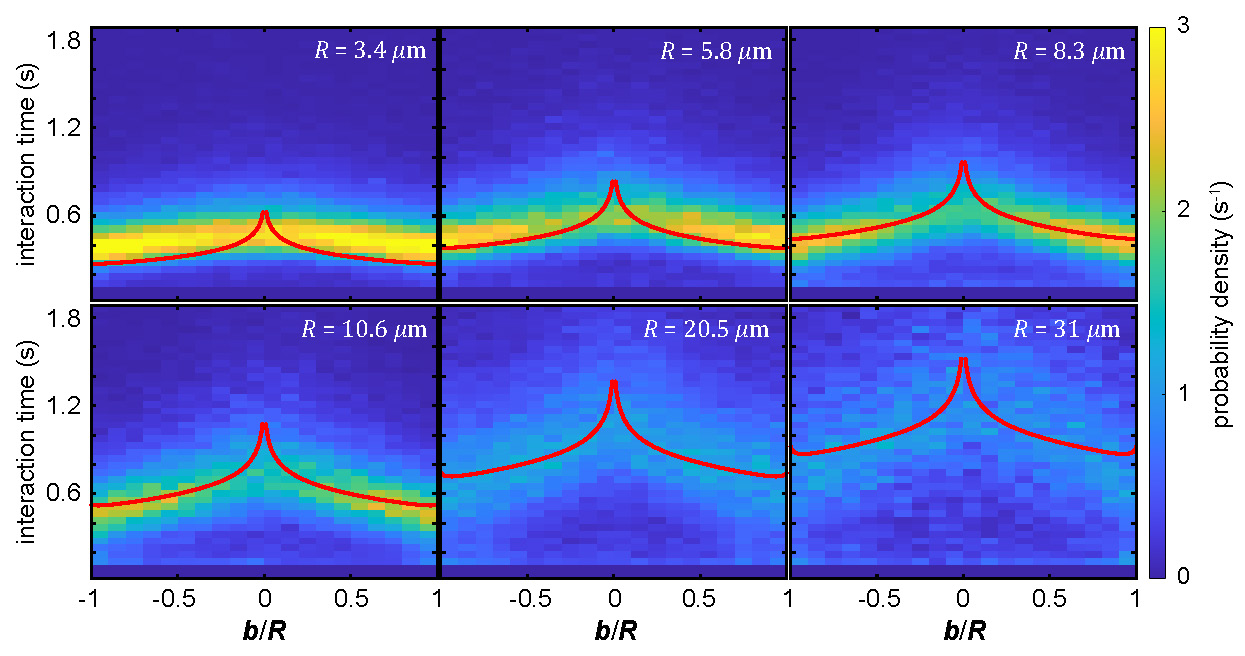}
\caption{Interaction time distributions as a function of dimensionless impact parameter $b/R$, across a range of pillar radii. The red lines show the model predictions, which were calculated by adding: (i) the transit time from interaction zone entry to pillar contact using the average cell speed, (ii) the time spent in contact with the pillar using integration of the differential equation, and (iii) the transit time from tangency to exiting the interaction zone using the average cell speed. Rotational diffusion shortens the sliding time as trajectories approach $b/R\rightarrow 0$, in a way that is not accounted for in the sterics-only model. Notably, the `signal-to-noise' ratio of measured data decreases with increasing pillar radius because the the number of pillars and hence number of interactions we can observe in a single field-of-view decreases faster than $R^{-2}$.}
\label{figinttime}
\end{center}
\end{figure}

\begin{figure}
\begin{center}
\includegraphics[width=4in]{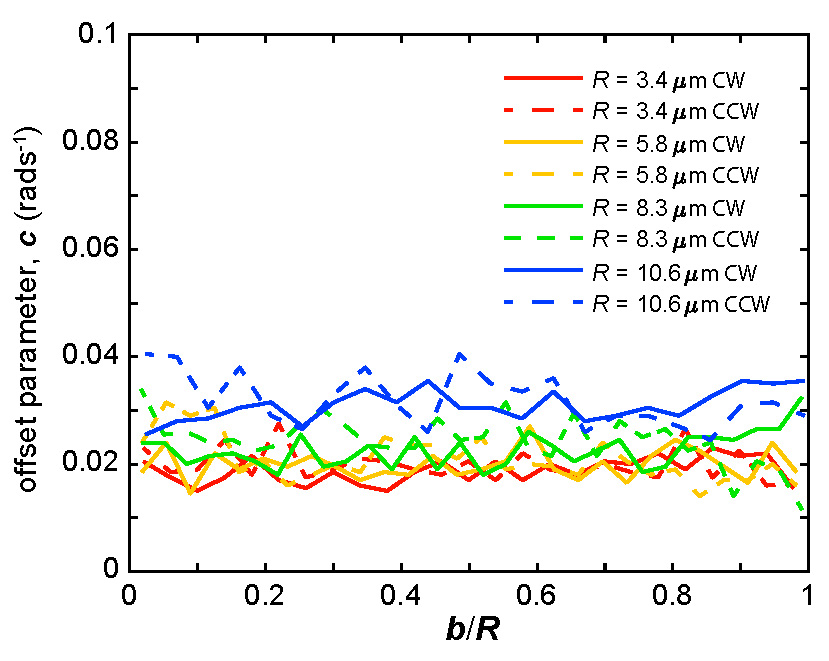}
\caption{Plot of the von Mises offset parameter (called $c$ above) as a function of $b/R$ across the four smallest radii.  The data are the modes from the MLE fits for the parameter estimation. The offsets are roughly constant across $|b/R|$ and approximately chirally symmetric, indicating that the frequency of random scattering events is independent of $|b/R|$ and not related to direction. There is also a rough upward trend in the offset with increasing pillar radius, indicating that random scattering is more common around larger pillars.  This may be related to the fact that larger pillars correspond to longer interaction times, and hence a higher probability of a random event (e.g. chemotactic tumble) during the interaction. It may also result from increased hydrodynamic trapping at larger radii, which causes cells to follow trajectories around the pillar for much longer times than steric scattering, but with a random detachment time, and hence random angle.}
\end{center}
\end{figure}

\begin{figure}
\begin{center}
\includegraphics[width=7in]{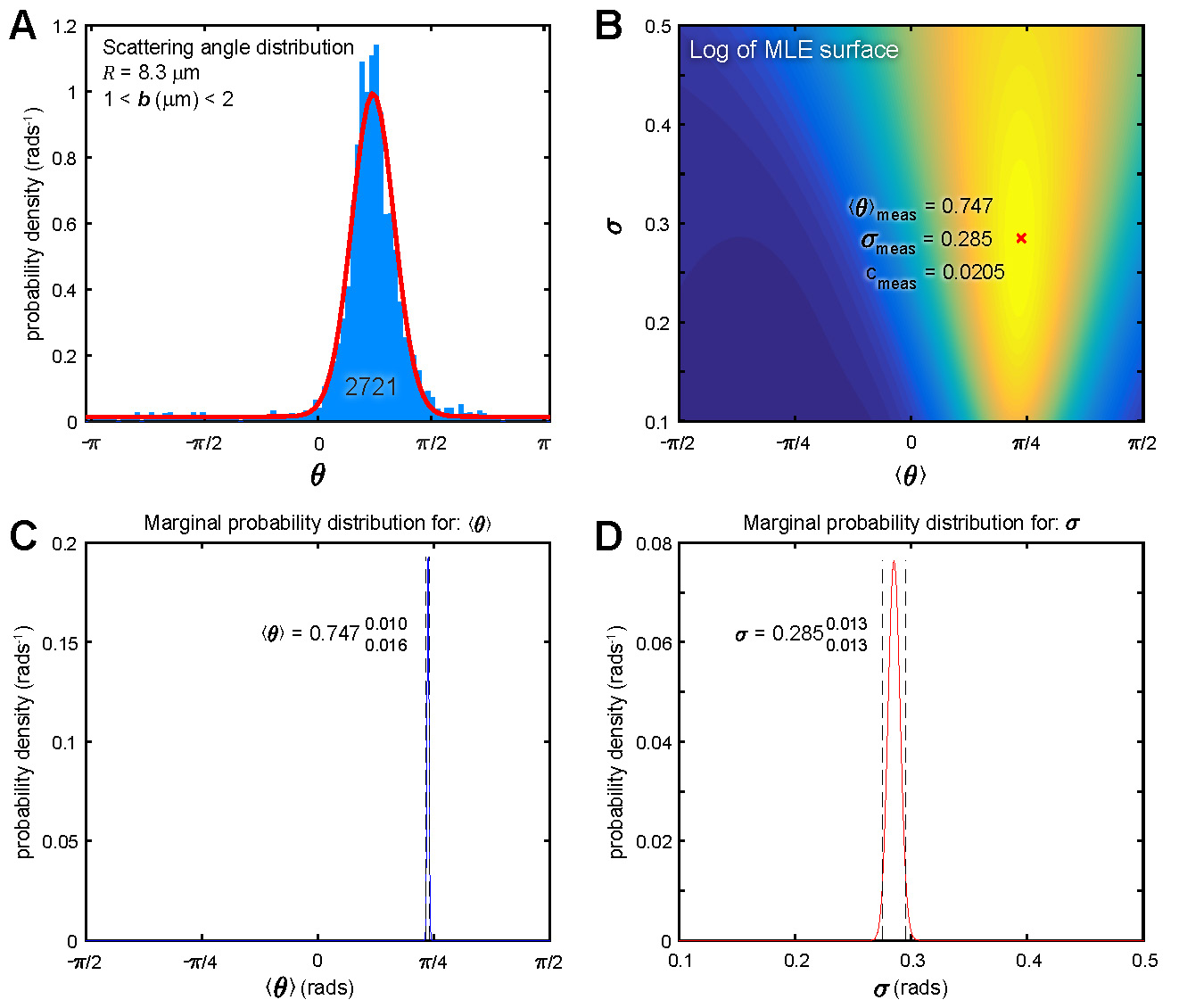}
\caption{Example output of the MLE fitting.  (A) A CW chiral scattering distribution with the MLE fit in red. (B) The natural log of the MLE fit surface for all data in the histogram, showing the mode values for all fit parameters. (C) The probability distribution for the measured value of $\left<\theta\right>$ showing the mode and 95\% confidence interval.  (D) The probability distribution for the measured value of $\sigma$  -- the width of the scattering distribution -- showing the mode and 95\% confidence interval.}
\label{fig_mle_fit}
\end{center}
\end{figure}

\begin{figure}
\begin{center}
\includegraphics[width=6in]{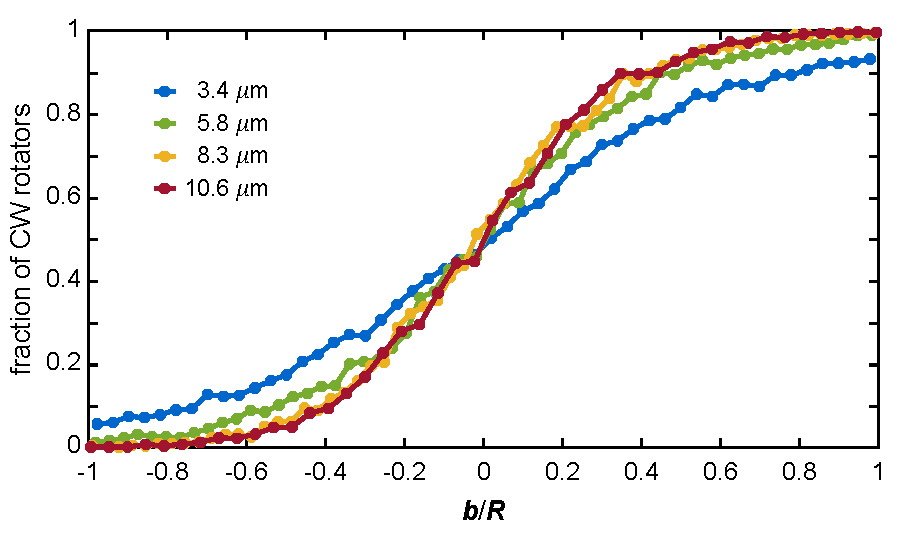}
\caption{Fraction of cells that rotate clockwise around a pillar as a function of dimensionless impact parameter. Assuming the pillar is centered on a local Cartesian coordinate system, clockwise rotation was defined by cell trajectories that crossed the center-line ($x = 0$) with $y>0$ in the rotated frame. The naive expectation from the steric model is that this would be an increasing step-function at $b/R=0$. Based on visual inspection of imaging data, as well as quantitative analysis of breaking the model assumption that the initial contact angle ($\alpha_o$) is set purely by $b$ and $R$, we hypothesize that fluctuations in cell orientation upon impact are what produce trajectories that traverse the pillar the `long way' around (i.e. opposite to the chirality predicted by the steric model). Such fluctuations are caused by translational and rotational diffusion of the cell body, as well as variations in cell morphology that affect initial contact angle.  If those fluctuations in orientation due to diffusion and morphology are rotationally isotropic, then we expect (and observe) that these curves are symmetric upon flipping about $b/R=0$ and $p_{CW}=1/2$, regardless of radius. }
\end{center}
\end{figure}

\begin{figure}
\begin{center}
\includegraphics[width=4in]{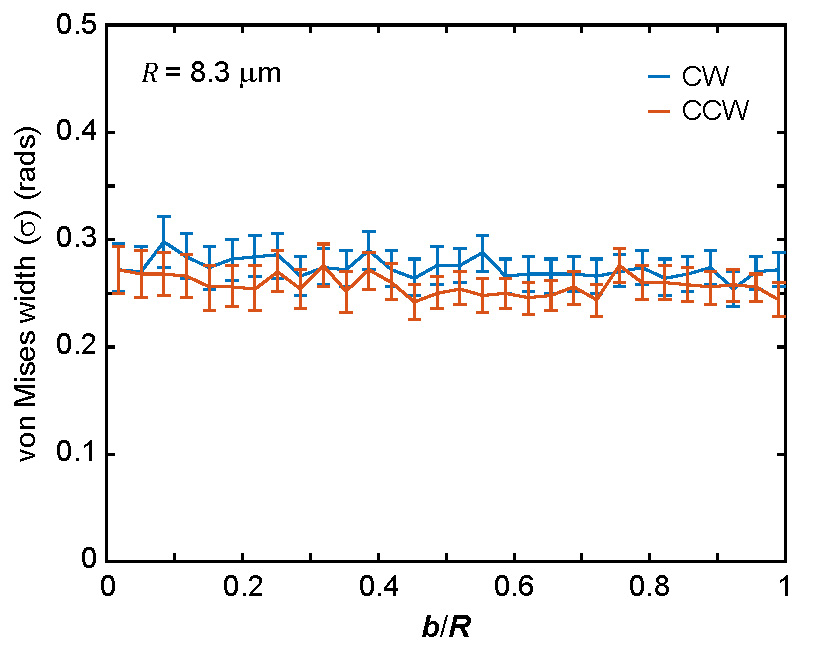}
\caption{Plot of the von Mises width parameter (called $\sigma$ above) as a function of $b/R$ for $R=8.3\,\mu m$.  The data are the modes from the MLE fits and the bounds are 95\% confidence intervals on the parameter estimation. The width parameter is approximately constant across all values of $b/R$ and is approximately chirally symmetric.}
\end{center}
\end{figure}

\begin{figure}
\begin{center}
\includegraphics[width=6in]{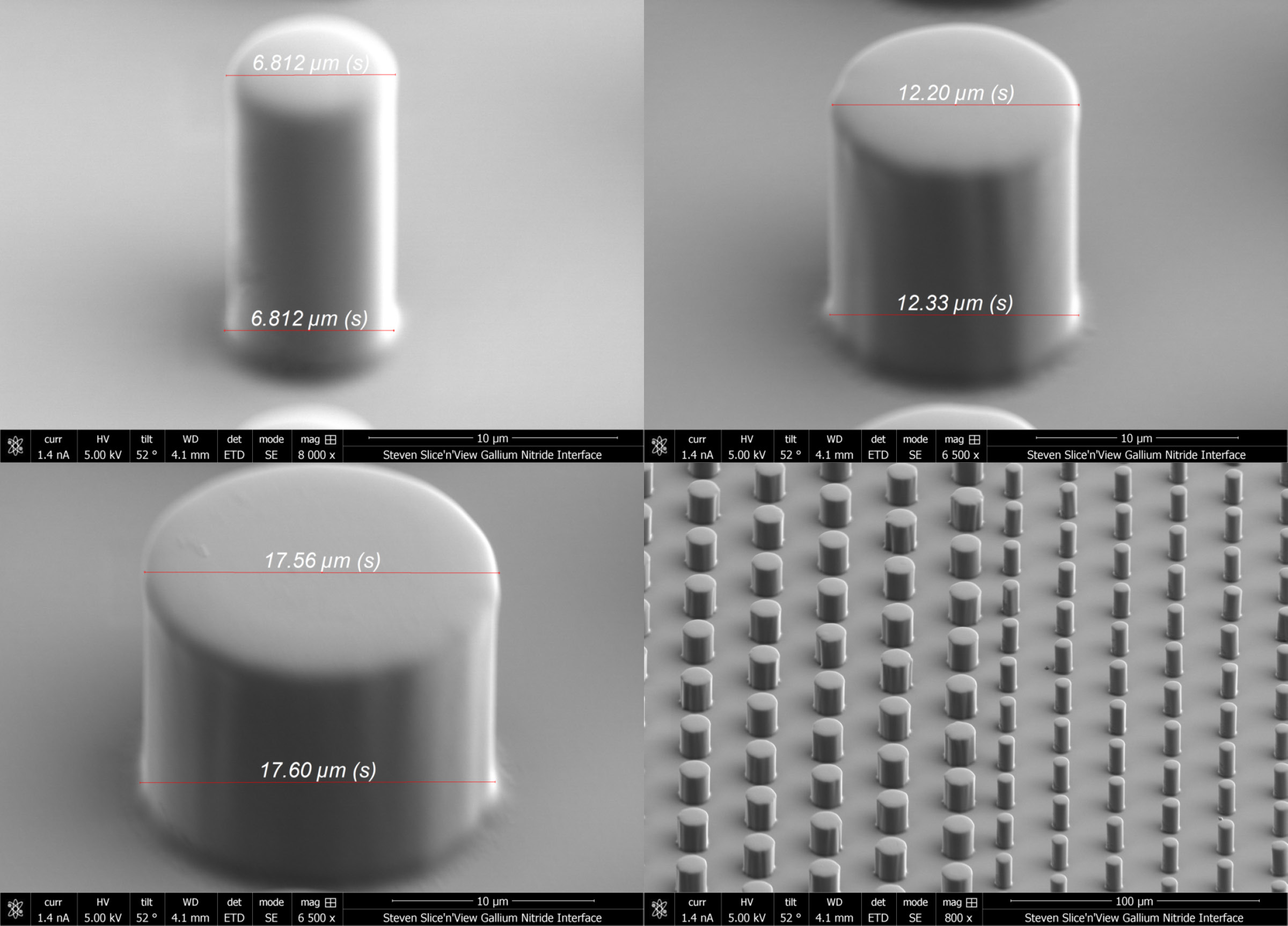}
\caption{Electron microscopy (EM) images of typical SU-8 polymeric pillars within our microfluidic devices. Pillar radii for each device region were measured using EM imaging. }
\label{fig_pillar_em}
\end{center}
\end{figure}

\begin{figure}
\begin{center}
\includegraphics[width=7in]{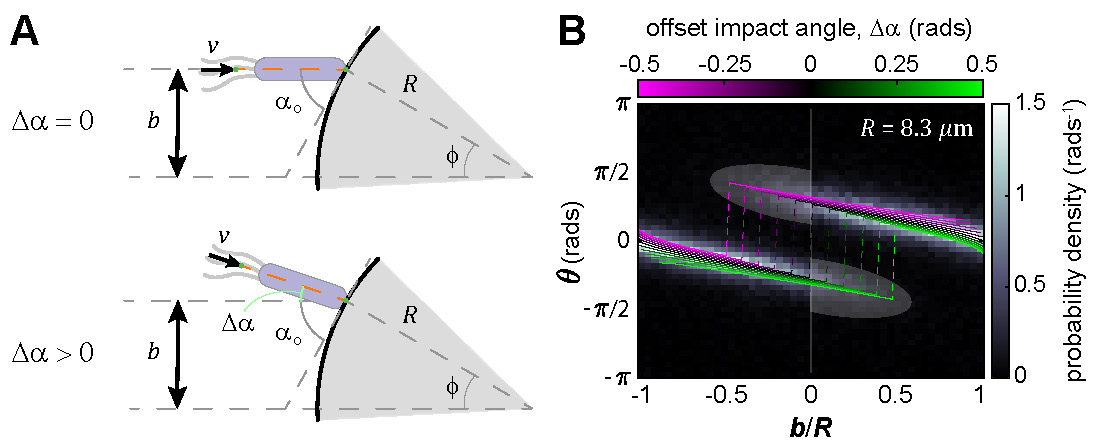}
\caption{Sensitivity of chirality and scattering angle on initial impact angle $\alpha_o+\Delta \alpha$. (A) Schematic showing the definition of the offset impact angle ($\Delta \alpha$). The model fits shown in the main text and preceding SI figures \ref{fig_theta_all_radii}, \ref{fig_beta_all_radii}, and \ref{figinttime} assume $\Delta \alpha=0$, in other words that $b$ and $R$ are the only parameters needed to determine $\alpha_o$.  However, all of our scattering data showed trajectories that circumvented the pillar the `long way' around, that is, with a chirality opposite to what is predicted by the steric model -- these are the highlighted lobes in (B).  We hypothesized that a combination of rotational diffusion and asymmetries in cellular morphology could lead to significant rotation of the cell body between entry into the interaction zone (which defines $b$) and contact with the pillar (which defines $\alpha_o$). We accounted for this possibility in the model by adding a constant offset ($\Delta \alpha$) to the initial impact angle ($\alpha_o$), and then calculated the resulting scattering angle $\left<\theta\right>$.  (B) As an example, we compare these scattering angle functions over a uniform range of offset impact angles ($\Delta \alpha$) (see colored lines and legend) to the measured data for $R=8.3\,\mu m$.  We found that (i) the lobes of measured, atypical chiral probability could be explained by reasonable values of $\Delta \alpha$, and (ii) that the observed spread in measured scattering angle for a particular value of $b/R$ could result from the same variations in $\Delta \alpha$.  Likewise, varying $\Delta \alpha$ also shifts the discontinuity (dashed vertical lines) along the $b/R$ axis in a way consistent with the observed probability distributions.}
\end{center}
\end{figure}

\begin{figure}
\begin{center}
\includegraphics[width=7in]{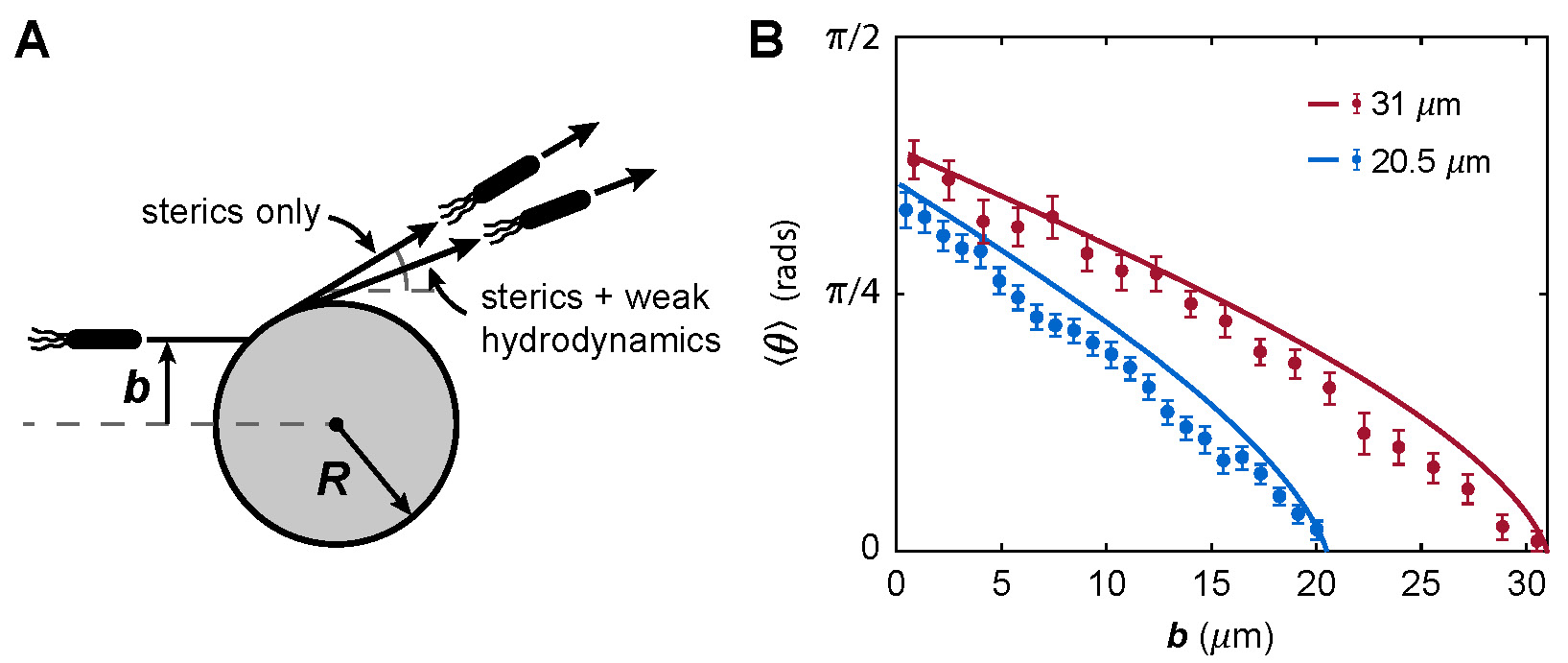}
\caption{(A) Schematic showing the relative scattering angles of a sterics-only scattering event vs. a scattering mechanism that involves hydrodynamic forces that attract the cell to the pillar surface and hence `over-rotate' it relative to the steric model. (B) Comparison of the model predictions (solid lines) to the measured data for mean scattering angle with 95\% confidence intervals around the mean, for the two largest pillars measured. The model overestimates the mean scattering angle at these larger radii, consistent with hydrodynamic forces near these low curvature surfaces over-rotating the cell relative to a sterics-only mechanism, and thus causing a smaller scattering angle.}
\end{center}
\end{figure}

\begin{figure}
\begin{center}
\includegraphics[width=7in]{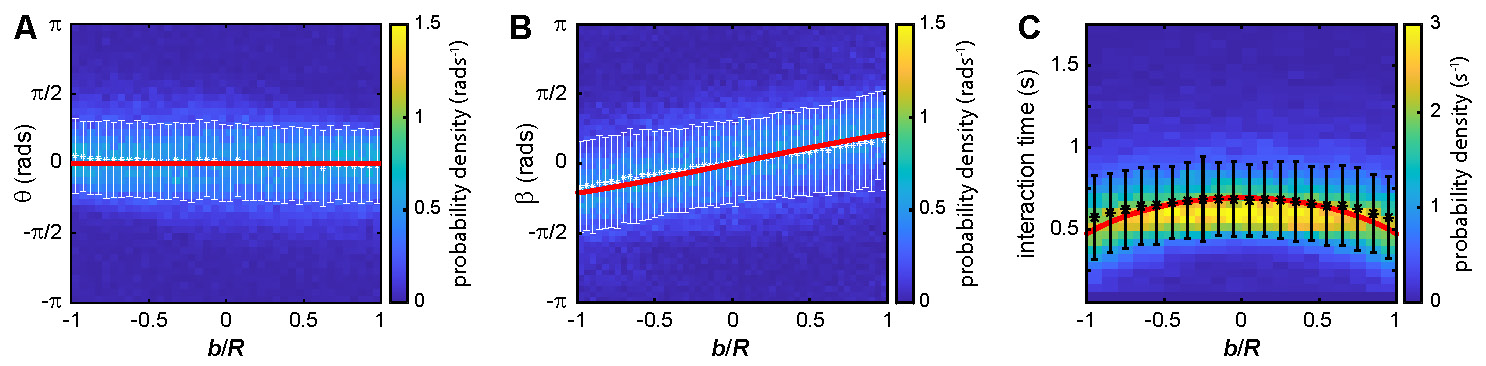}
\caption{Comparison of data and null-model predictions in the case of no steric interaction. We collected imaging data in a featureless area of our microfluidic device and calculated the same relationships for scattering angle (A, $\theta$), exit angle (B, $\beta$), and interaction time (C), assuming a nominal fictitious pillar size of $R=5.8\,\mu m$ with an interaction zone of $\delta = 2.2\,\mu m$.  We used the full data collection and analysis pipeline employed with `real' steric interaction data to this scenario that lacked steric interactions (call this the `null model'). The null model makes specific, quantitative predictions of the (mean) relationships between dimensionless impact parameter ($b/R$) and, respectively, scattering angle ($\theta$), exit angle ($\beta$), and interaction time.  The heat maps are the measured control data, the red lines are the zero-fit predictions of the null model, again assuming the same $L=3.75\,\mu m$. The points (white in A and B, black in C) are the means of the measured control data suitable for comparison to the null model.  Note that the predictions for $\left<\theta\right>$ and $\left<\beta\right>$ under the null model are starkly, qualitatively distinct from the predictions of the steric model. These mean values show a mild systematic deviation from the null model as $|b/R|\rightarrow 1$ that lies within a standard deviation of the mean of the data (vertical data bars). We speculate that this results from differences in path length and number-density of paths exiting the interaction zone along its circular boundary.  Such deviations break the null-model assumption of persistence length $\lambda\gg(R+\delta)$, producing an asymmetry that progressively grows as $|b/R|$ increases. }
\label{figcontrol}
\end{center}
\end{figure}

\begin{figure}
\begin{center}
\includegraphics[width=7in]{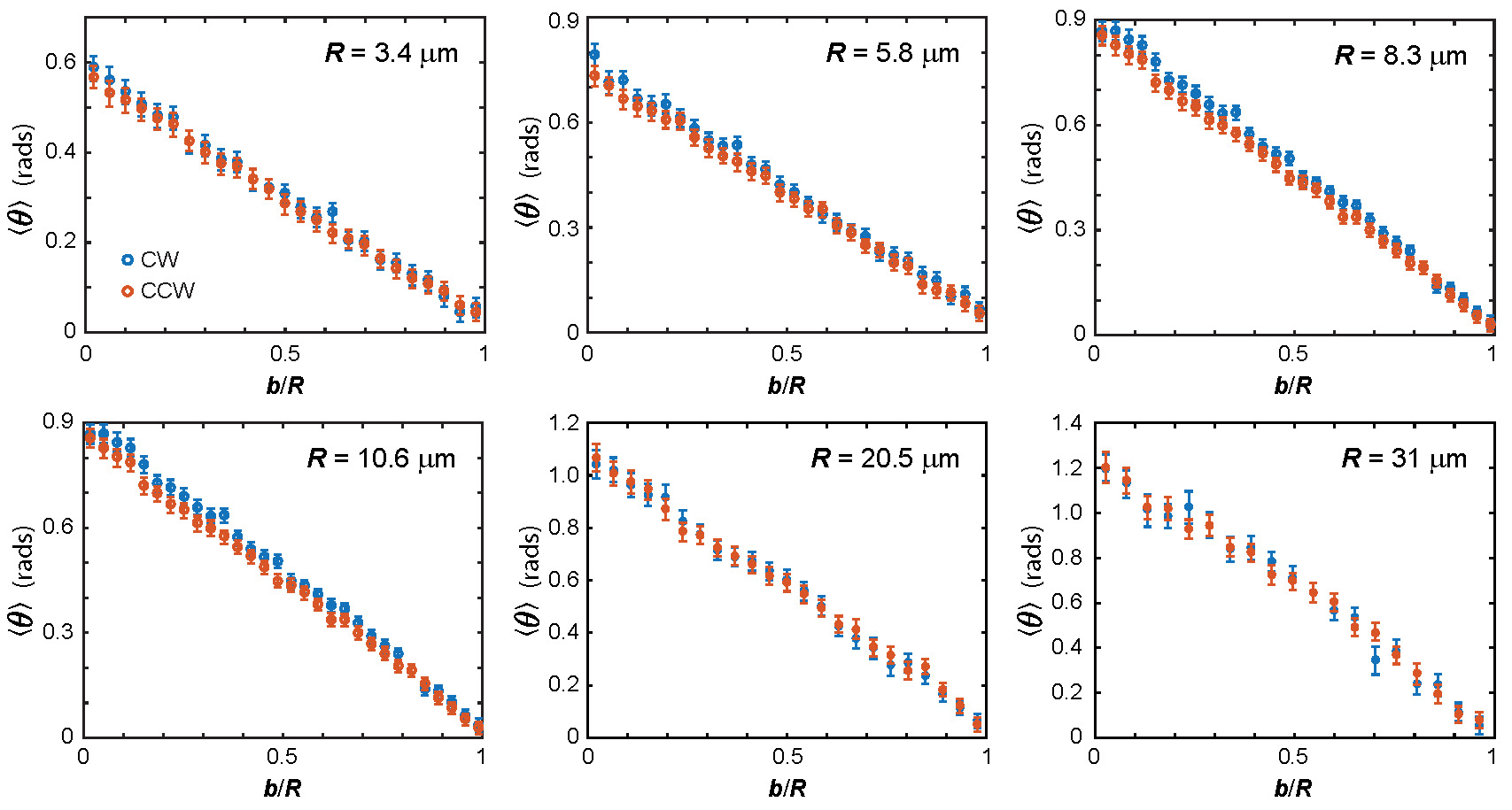}
\caption{Based on the symmetries present in the propulsion of the bacteria and within the microfluidic device, the distribution of scattering angles as a function of dimensionless impact parameter should be -- regardless of mechanism -- symmetric when mirrored about both the $\theta=0$ and $b/R=0$ axes.  Using the MLE fits to a modified von Mises distribution, here we plot $\left<\theta\right>$ vs. $b/R$ with 95\% confidence intervals, with the appropriate mirroring to plot the CW and CCW trajectories overlaid. Across the range of $b/R$, the data appear approximately symmetric, with mild systematic asymmetry for some radii. These slight chiral asymmetries are likely due to a combination of (observed) systematic asymmetries in the radius of the pillars with height due the fabrication process (see Fabrication Details and electron microscopy images, SI Fig.~\ref{fig_pillar_em}.}
\label{fig_chiral_test}
\end{center}
\end{figure}

\end{document}